\documentclass[preprintnumbers,amsmath,amssymb,floatfix,11pt,prd,onecolumn,superscriptaddress,nofootinbib]{revtex4}
\usepackage[utf8]{inputenc}
\usepackage{diagbox}
\usepackage{latexsym}
\usepackage{epsfig}
\usepackage{epstopdf,float}
\usepackage{graphicx}
\usepackage{amssymb}
\usepackage{amsmath}
\usepackage{dcolumn}
\usepackage{bm}
\usepackage{color}
\usepackage{comment}
\usepackage[shortlabels]{enumitem}
\usepackage{subfigure}
\usepackage{multirow}
\usepackage{xcolor}
\begin{document}

\title{\bf Magnetic Reconnection and Energy Extraction from a Rotating Black Hole in the Einstein-AdS SU(N)-Nonlinear Sigma Model}

\author{Muhammad Israr Aslam}
\email{mrisraraslam@gmail.com, israr.aslam@umt.edu.pk}
\affiliation{Department of Mathematics, School of Science, University of Management and Technology, Lahore-$54770$, Pakistan.}

\author{Muhammad Nawaz}
\email{mnawazzz158@gmail.com}
\affiliation{Department of Mathematics, University of Okara, Okara-56300 Pakistan}

\author{Abdul Malik Sultan}
\email{ams@uo.edu.pk; maliksultan23@gmail.com}
\affiliation{Department of Mathematics, University of Okara, Okara-56300 Pakistan}

\author{Ke Wang}
\email{kkwwang2025@163.com}\affiliation{School of Material Science and Engineering, Chongqing Jiaotong University, Chongqing 400074, China}

\author{Hamood Ur Rehman}
\email{hamood@uo.edu.pk}
\affiliation{Department of Mathematics, University of Okara, Okara-56300 Pakistan}

\author{Yakup Yildirim}
\email{yyildirim@biruni.edu.tr}
\affiliation{Department of Computer Engineering, Biruni University, Istanbul–34010, Turkey.}

\begin{abstract}
In the present study, we analyze the power and efficiency of energy extraction via magnetic reconnection in the rotating Einstein-AdS-SU($N$)-NLSM black hole, both in the circular-orbit
regime and in the plunging region. Initially, we define the background properties of this spacetime, and then analyze the physical quantities such as the size of the ergoregion, the event
horizon, and the boundaries of the ergosphere. We analyze the magnetic reconnection process within circular orbits. We plot energy-extraction parameter diagrams and analyze the power and
efficiency of energy extraction. Our results indicate that energy extraction remains feasible even at a spin parameter as low as $0.7$, significantly below previously reported thresholds, and the extracted power can exceed that of the Blandford-Znajek mechanism with specific constraints. The coupling constant $K$, AdS radius $l$ and the flavors number $N$, collectively participate in lowering the spin threshold for energy extraction. Consequently, we further investigate the permissible energy extraction region for the energy extraction mechanism in the plunging region, as well as the corresponding power output and efficiency. We observe that the energy extraction is possible even at a spin as low as $0.2$. Importantly, the parameters $N,~K$ and $l$ mainly contribute to lowering the energy extraction spin threshold. This behavior is similar to that in circular orbits. Finally, comparing the plunging region with the circular orbits, we observe that the energy extraction power in the plunging region is higher than in the circular orbits.
\end{abstract}
\maketitle

\section{Introduction}
Black holes are among the most fascinating and enigmatic objects in the universe. These extraordinary celestial bodies were predicted by Albert Einstein through his theory of general relativity and are known for their immense gravitational pull, from which not even light can escape. They play a fundamental role in astrophysics and celestial mechanics, influencing the motion of stars, galaxies, and surrounding matter. Over the past few decades, numerous astronomical observations and discoveries, including detection of gravitational waves and specifically the image released by the Event Horizon Telescope (EHT), have provided strong evidence for their existence, making black holes one of the most important subjects in modern theoretical and observational physics. Black holes formed by the natural consequences of general relativity \cite{penrose1965gravitational}. And they have been part of the efforts being made to reconcile general relativity with quantum mechanics. In the fabric of general relativity, the no-hair theorem is a foundational principle in black hole physics, which state that any asymptotically flat, stationary, axially symmetric solution of a black hole to the Einstein field equations can be fully described by three parameters: mass $M$, electric charge $Q$ and  angular momentum $j$ \cite{carter1971axisymmetric}, which lie in the Kerr-Newman solution \cite{newman1965metric}. In the fabric of general relativity, the no-hair theorem is a foundamental principle in black hole physics, which states that the Kerr-Newman metric is the sole solution to the Einstein-Maxwell equations that satisfy the conditions of axial symmetry and stationarity.  Black holes are often considered electrically neutral for practical applications, because any residual charge is quickly neutralized through interactions with surrounding matter \cite{israel1968event}.

There has been considerable debate regarding the mathematical foundation of the no-hair theorem, particularly concerning the assumption of analyticity. The assumption that, smooth analytic metrics are perfect representations of black holes is crucial to the theorem, but it may not be universally applicable in all theoretical frameworks. Modifications to general relativity, for example, which are found in metric affine gravity theories, allow the black holes that different from the Kerr solution. Testing the no-hair theorem thus involves exploring possible deviations from Kerr metrics, particularly in scenarios proposed by extended theories of gravity. Subsequently, the rotating black hole metric is the standard model for describing astrophysical black holes and their observable properties. Both observational and theoretical efforts are required to disprove the existence of non-Kerr black holes \cite{ryan1995gravitational}.

At the same time, both in theoretical and phenomenological terms, the nonlinear sigma model (NLSM) has become a crucial effective field theory, which has widespread applications in quantum field theory, string theory and statistical mechanics \cite{new1}.  The NLSM is used to describe the low-energy dynamics of pions \cite{nair2006quantum},  typically in the fabric of the internal symmetry group SU($2$), which represents the two-flavour case. While the predictions made by the NLSM align well with experimental results, solving the field equations derived from the model can be complex, as these equations are generally nonlinear and coupled, involving ($N^2-1$) equations where $N$ is the flavor number encoded in the SU$(N)$ group. Consequently, there are many solutions that have been constructed numerically through effective implementations. In this field, some of the most important results are presented in Refs. \cite{nelmes2011skyrmion,luckock1986black}. The corresponding equations of motion become more complex, when the NLSM is coupled with general relativity or Maxwell theory to investigate the more intricate physical systems.  Nonetheless, in \cite{canfora2013nonlinear}, the authors proposed a methodology that go beyond spherical symmetry, which have been used to find the exact solutions in  Skyrme model, and the Yang-Mills-Higgs theory. These solutions describe a variety of objects that include crystalline structures of topological solitons \cite{canfora2018ordered},  topological solitons at finite volume \cite{alvarez2017integrability}, gravitating solitons \cite{ayon2016analytic},  black strings \cite{astorino2018black}, black holes \cite{astorino2018hairy}, and boson stars \cite{giacomini2018solitons}.

There is a large amount of extractable energy in the vicinity of a rotating black hole, which is predicted by general relativity. And the process of extracting energy from a rotating black hole has been a very interesting topic among the scientific community. The Penrose process \cite{wald1974energy} represents the first proposed mechanism for extracting energy from a rotating black hole. It operates within the ergo-sphere, a region where negative energy orbits can exist, enabling the extraction of rotational energy. Black hole attracts a particle of energy $E$ from infinity. And this particle splits into two parts within the ergo-sphere. One of them carries negative energy and crosses the event horizon, while the other carries positive energy and escapes from the event horizon; thereby enabling energy extraction. However, the Penrose process is an ideal framework. Building on this mechanism, numerous scientists have proposed a variety of alternative energy extraction processes that extend and refine the original concept. (for instance one can see Refs.
\cite{teukolsky1974perturbations,piran1975high,blandford1977electromagnetic,takahashi1990magnetohydrodynamic}).

A fundamental process that occurs in plasma is magnetic reconnection, in which magnetic field lines break and reconnect into a new configuration, releasing a significant amount of energy. This
mechanism plays a crucial role in energy extraction from rotating black holes. Specifically, Comisso and Asenjo \cite{comisso2021magnetic} demonstrated that, in the vicinity of a rotating black hole, the power extracted via magnetic reconnection can exceed that of the usual Blandford-Znajek mechanism \cite{blandford1977electromagnetic}. After that, this research was rapidly enlarged to incorporate the other models of rotating black holes
\cite{wei2022effects,israr1,khodadi2022magnetic,carleo2022energy,wang2022extracting,li2023energy,l2023energy,zhang2024energy,zhang2024,khodad2023harvesting,shaymatov2024kerr,rodriguez2025energy,long2025magnetic,zeng2025energy,wang2025energy,zeng2025e2,eshtursunov2026energy,eshtursunov2026magnetic,yao2026energy,enhanced,yuchih2025energy,cheng2025extractin}. The spin of a black hole for the  Blandford-Znajek mechanism does not require a slow spin. The minimum spin of 0.85 is required for
the Kerr black hole for some fixed values of parameters $\xi=\pi/12$ and $\sigma=100$ \cite{comisso2021magnetic}. Under the same condition, the minimum spin is found to be no less than $0.7$ for the Kerr-de Sitter spacetime \cite{wang2022extracting}. The minimum spin is required about $0.79$  for a rotating regular black hole \cite{li2023energy}. For a rotating hairy black hole, the minimum spin is about 0.86, under the identical conditions \cite{l2023energy}. Overall, the results indicate that, across different gravitational backgrounds, the black hole spin plays a major role in determining the efficiency and conditions of magnetic reconnection. A particularly promising research direction is the study of energy extraction via magnetic reconnection at lower black hole spin values in alternative gravitational spacetimes, which has attracted considerable attention and continues to engage researchers
in the field.

Building upon these scenarios, we look into the magnetic reconnection in the neighborhood of the rotating Einstein-AdS-SU($N$)-NLSM (RASN) black hole  model \cite{fathi2025shadows}. Our objective is to examine how the spin parameter $a$, coupling constant $K$, AdS radius $l$ and the flavors number $N$ influence the magnetic reconnection process in the RASN black hole model, together with the corresponding power output and efficiency of energy extraction. In particular, we aim to determine whether this generalized rotating and accelerating black hole configuration provides a more effective mechanism for energy extraction compared to previously studied rotating black hole models. We demonstrate that energy extraction through magnetic reconnection remains feasible even at moderate spin values, indicating an enhanced performance of the system. This improvement can be attributed to the combined effects of acceleration and gravitomagnetic charge, which modify the spacetime structure and effectively reduce the critical spin threshold required for efficient energy extraction.

The remainder of this paper is organized as follows. In section {\bf II}, we briefly define the background of the RASN black hole model and discuss its geometrical and physical properties. In Section {\bf III. A}, we describe the magnetic reconnection process in circular orbits. In section {\bf III.B}, we analyzes the allowed parameter space for energy extraction. While section {\bf III.C} is devoted to discussing the corresponding power and efficiency of energy extraction. In section {\bf IV.A}, we present the magnetic reconnection mechanism in the plunging region, while in section {\bf IV.B}, we analyzes the associated power and efficiency of energy extraction. Finally, our conclusions and discussions are presented in the last section.

\section{A Brief Review of RASN Black Hole Model}

\begin{figure}[H]
\begin{center}
\subfigure[~$l=15,K=1/100,N=2$]{\includegraphics[width=4.7cm,height=4.3cm]{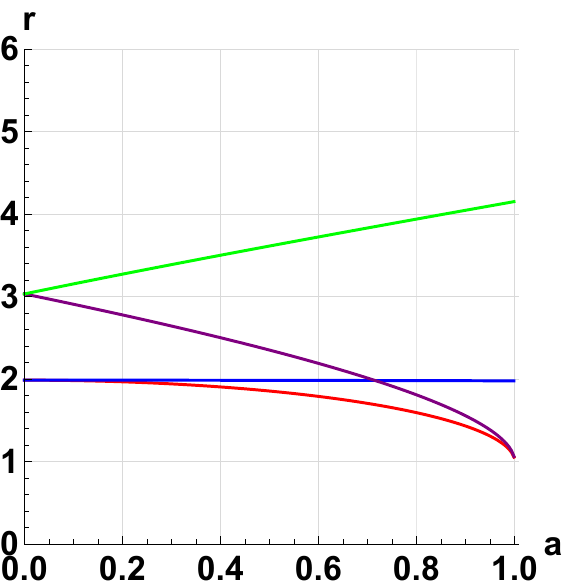}}
\subfigure[~$l=6,K=1/100,N=2$]{\includegraphics[width=4.7cm,height=4.3cm]{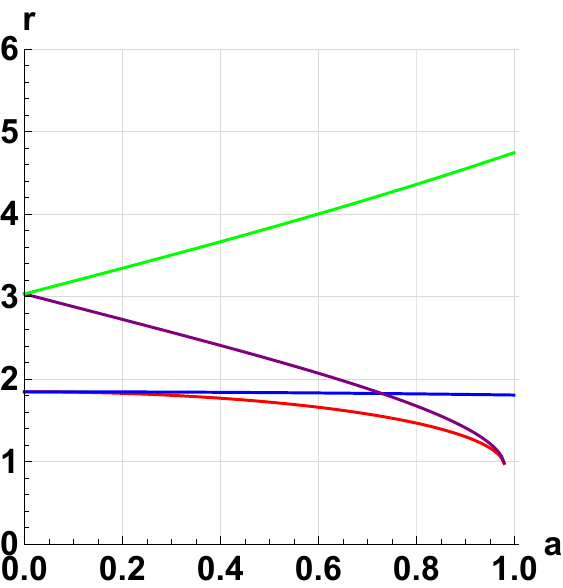}}
\subfigure[~$l=3,K=1/100,N=2$]{\includegraphics[width=4.7cm,height=4.3cm]{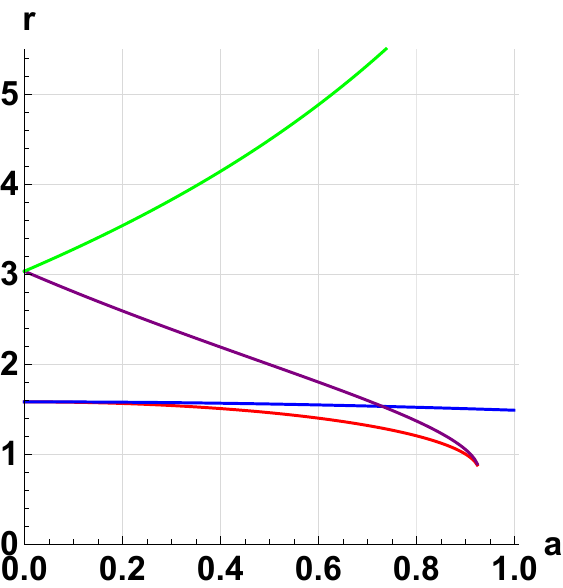}}
\subfigure[~$N=1,K=1/100,l=15$]
{\includegraphics[width=4.7cm,height=4.3cm]{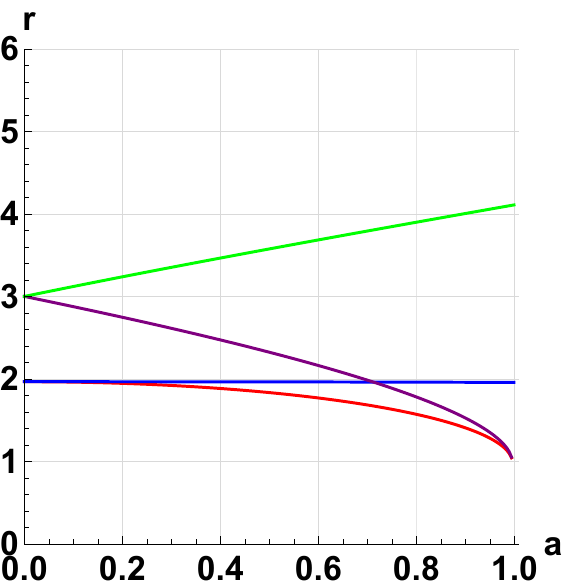}}
\subfigure[~$N=3,K=1/100,l=15$]{\includegraphics[width=4.7cm,height=4.3cm]{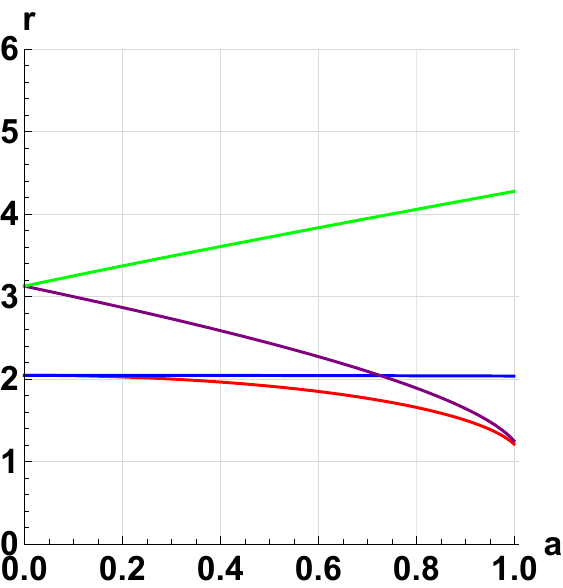}}
\subfigure[~$N=5,K=1/100,l=15$]{\includegraphics[width=4.7cm,height=4.3cm]{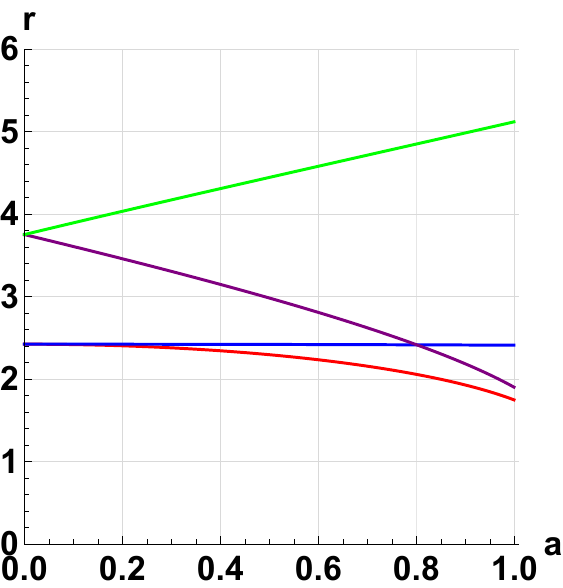}}
\subfigure[~$K=1/100,l=15,N=2$]
{\includegraphics[width=4.7cm,height=4.3cm]{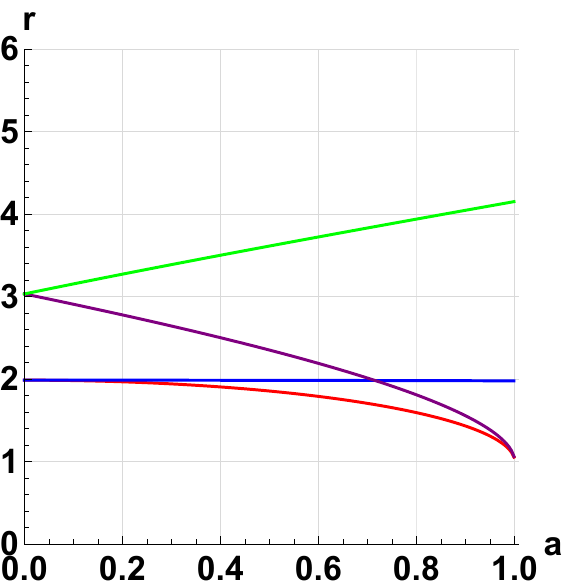}}
\subfigure[~$K=1/10,l=15,N=2$]{\includegraphics[width=4.7cm,height=4.3cm]{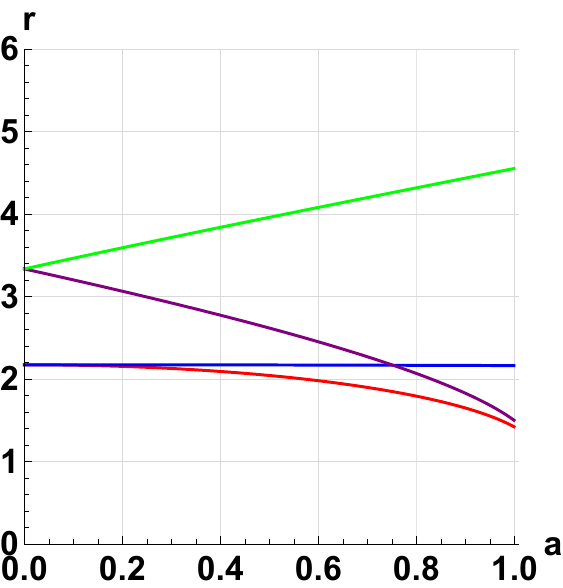}}
\subfigure[~$K=1/5,l=15,N=2$]{\includegraphics[width=4.7cm,height=4.3cm]{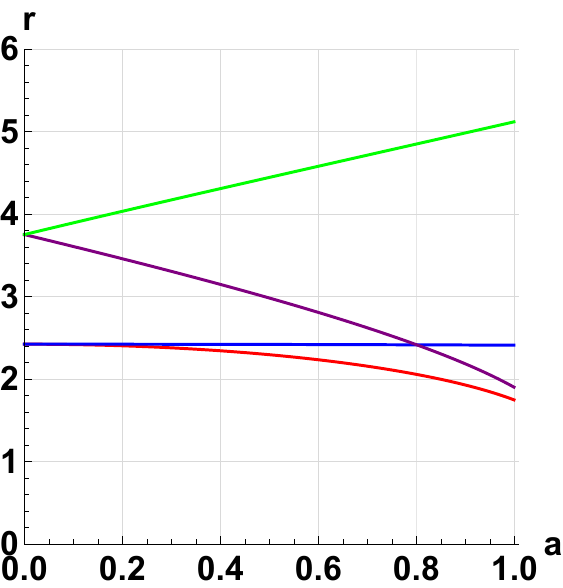}}
\caption{Plots showing the photon sphere radius, ergosphere boundary and event horizon with respect to $a$. The first row shows changes in the four curves for fixed $K$ and $N$ and varying $l$. The second row shows changes for fixed $l$ and $K$ and varying $N$. The third corresponds to fixed $l$ and $N$ and varying $K$. The green, purple, blue and red curves correspond to the retrograde orbits, prograde orbits, the boundary of the ergosphere, and the event horizon, respectively.}\label{fig1}
\end{center}
\end{figure}

An analysis of black hole solutions within the framework of the Einstein-SU($N$)-NLSM is essential for assessing the physical viability and consistency of the model. The Einstein-SU($N$)-NLSM is characterized by the action \cite{henriquez2022black}
\begin{equation}
\mathcal{S}=\int d^{4}x \sqrt{-g}\left(\frac{1}{2\kappa}(R-2\Lambda)+\frac{K}{4}\mathrm{Tr}[L^{\mu}L_{\mu}]\right),
\label{sun function}
\end{equation}
where $R$ denotes the Ricci scalar, $\Lambda$ represents the cosmological constant, and $L_{\mu}$ corresponds to the components of the Maurer-Cartan form, which is expressed as
\begin{equation}
L_{\mu}=U^{-1}\partial_{\mu}U=L_{\mu}^{i}t_{i}.
\label{maurar}
\end{equation}
Here, $U(x)\in \mathrm{SU}(N)$, with $N$ denoting the number of flavors embedded within the SU($N$) Lie group \cite{bertini2006euler,cacciatori2017compact,tilma2002generalized}, and $t_i$ being the generators of the SU($N$) Lie algebra, where $i=1,\ldots,(N^{2}-1)$. In this framework, $k$ is the gravitational constant, while $K$ is a positive coupling constant determined through experimental observations. The Einstein-SU($N$)-NLSM framework gives rise to deformed  Schwarzschild-AdS black holes \cite{henriquez2022black}, describing an asymptotically AdS static
and spherically symmetric black hole spacetime. The corresponding line element is given by
\begin{equation}
ds^{2}=-f(r)dt^{2}+\frac{1}{f(r)}dr^{2}+r^{2}d\theta^{2}+r^{2}\sin^{2}\theta d\phi^{2},
\label{lineelement}
\end{equation}
The main characteristics of this class of black holes are twofold: first, they are supported by pionic matter, and second, they extend the properties of SU($N$) black holes within a broader theoretical framework \cite{canfora2013hedgehog,gibbons2005self}. Notably, this black hole solution exhibits an asymptotic structure corresponding to the AdS version of the Barriola-Vilenkin metric \cite{rhie1991global} and reduces to the Schwarzschild-AdS spacetime in the limit $K=0$. The Kretschmann scalar is explicitly given by
\begin{equation}
\begin{aligned} \mathcal{K}
&=R^{\xi\zeta\alpha\beta}R_{\xi\zeta\alpha\beta}\\
&=\frac{1}{ 9r^{6} }
\Big[{ k^{2}K^{2}(N^{3}-N)^{2}r^{2} +24k KMN(N^{2}-1)r +4k K\Lambda N(N^{2}-1)r^{4} +432M^{2} +24\Lambda^{2}r^{6} }
\Big],
\label{Kretschmann}
\end{aligned}
\end{equation}
and the corresponding lapse function $f(r)$ is defined as
\begin{equation}
f(r)=1-\frac{2M}{r}-\frac{\Lambda}{3}r^{2}-Kk a_{N}.
\label{lapse}
\end{equation}
Here, $M$ represents the black hole mass, while the positive quantity $a_{N}$ is given by
\begin{equation}
a_{N}=\frac{N(N^{2}-1)}{6}.
\label{positiveqquantity}
\end{equation}
One important feature of the lapse function (\ref{lapse}) is the emergence of a deficit angle, which is explicitly encoded in the term $Kka_{N}$. Such a feature is commonly encountered in black hole spacetimes sourced by a cloud of strings (see, for example see Refs.~\cite{letelier1979clouds,toledo2018black,gracca2018cloud}). It has also been shown that the presence of a deficit angle can significantly affect the motion of both massive and massless test
particles, as it enters directly into the corresponding equations of motion \cite{batool2017null,mustafa2021radial,fathi2022study}. A straightforward analysis of the Kretschmann scalar with $K=0$ and $\Lambda=0$ shows that it reduces to $48M^{2}/r^{6}$, which
corresponds to the Kretschmann scalar of the Schwarzschild black hole. In this work, we use Boyer-Lindquist (BL) coordinates and geometric unit (G=c=1). Finally, the RASN black hole model is
defined as \cite{fathi2025shadows}
\begin{equation}
\begin{aligned}
ds^{2} = &\;
\frac{\Delta_{\theta}\sin^{2}\theta}{\Sigma^{2}\rho^{2}}
\left(a\,dt-(r^{2}+a^{2})\,d\phi\right)^{2} & +
\frac{\rho^{2}}{\Delta_{r}}\,dr^{2} +
\frac{\rho^{2}}{\Delta_{\theta}}\,d\theta^{2} & -
\frac{\Delta_{r}}{\Sigma^{2}\rho^{2}} \left(dt -
a\sin^{2}\theta\,d\phi\right)^{2},
\end{aligned}
\label{metric}
\end{equation}
where
\begin{equation}
\Delta_{r}=(r^{2}+a^{2})(1-\frac{\Lambda}{3}r^{2})-2Mr-\frac{1}{6}kKN(N^{2}-1)r^{2},
\label{deltaR}
\end{equation}
\begin{equation}
 \rho^{2}=r^{2}+a^{2}\cos{\theta} , \quad \Delta_{\theta}=1+\frac{a^{2}}{3}\Lambda\cos^{2}{\theta},\quad
 \Sigma=1+\frac{\Lambda}{3}a^{2}, \quad
 \Lambda=-\frac{3}{l^{2}},
\label{metricfunction}
\end{equation}
in which $a$ denotes the black hole  spin parameter, $l$ is AdS curvature radius, while the parameters $K$, $N$, and $k$ characterize the deviations of the RASN black hole from the standard Kerr-AdS black hole spacetime and are directly linked to the deficit angle. If we put the value of one of them equal to zero $K=N=k=0$ and with one special case $N=1$, then the RASN black hole metric is reduced to Kerr-AdS spacetime. The horizons of the RASN black hole are determined by $\Delta_{r}=0$. This equation has two positive roots, one of the largest is called event horizon $r_{+}$ and the smaller one is called the inner (Cauchy) horizon $r_{-}$. Also, the boundary of the ergosphere is determined by $g_{tt}=0$, which has one root outside the event horizon. On the equatorial plane ($\theta=\pi/2$), we consider the equation of motion for particles with $M=1$. From the Hamilton-Jacobi equation, one can obtain the radial equation as
\begin{align}
\left(\frac{dr}{d\lambda}\right)^2
= \Sigma^2[(r^{2}+a^{2})E - aL]^2
- \Delta_r\left[(L-aE)^2 + \mathcal{O}\right]= R(r),
\label{Rfunction}
\end{align}
where $\mathcal{O}$ is Carter constant \cite{carter1968global} satisfying $\mathcal{O}=\mathcal{Q}-(L-aE)^2$ and $\mathcal{Q}$ is the Carter's separation constant, $E$ is the energy, $L$ is the angular momentum, $\lambda$ is the Mino time \cite{mino2003perturbative}, related to proper time $\tau$ by $\frac{d\tau}{d\lambda} = \rho^2$. The particles we examine are considered to be in circular orbits in the equatorial plane. And the radii of these circular orbits range from infinity down to the photon sphere. The photon sphere radius satisfies
\begin{equation}
R(r)=0, \qquad \frac{dR(r)}{dr}=0.
\label{conditionforR }
\end{equation}
Generally, these conditions give two solutions, in which one of them is for prograde orbits, and the other one is for retrograde orbits. Initially, we plot the photon sphere radius, boundary of ergosphere, and event horizon $r$ with respect to $a$ as shown in  Fig. \ref{fig1}. The green, purple, blue and red curves represent to the retrograde orbits, prograde orbits, the boundary of the ergosphere, and the event horizon, respectively. From these graphs, we found that the boundary of the ergosphere is smaller than $2$, and the maximum allowed spin for RASN black hole is less than $1$. In the first row of Fig. \ref{fig1}, for the fixed values of $K$ and $N $, as $l$ decreases, the maximum allowed spin is also decrease. In the second and third rows, we observe that for fixed values of $K$ and $l$, an increase in $N$ leads to a similar behavior in the event horizon, photon sphere radius, and the boundary of the ergosphere. Subsequently, in the third row, for fixed values of $l$ and $N$, increasing $K$ produces the same qualitative behavior in these quantities. Furthermore, the parameters $N$ and $K$, together with decreasing $l$, enhance the deviation from the RASN black hole. The difference is that, in the former case, where $N$ and $K$ increase, the positions of the event horizon, photon sphere radius, and ergosphere boundary all increase, while the ergosphere boundary does not exhibit any noticeable bending. In contrast, in the latter case, where $l$ decreases, the positions of the event horizon and ergosphere boundary decrease, and the ergosphere boundary becomes bent or distorted. Since energy extraction occurs between the boundary of ergosphere and the event horizon, so we only need to consider prograde
orbits (see Fig. \ref{fig1}). At the photon sphere radius, the particle's energy and angular momentum diverge to infinity. Therefore, for circular orbits, the energy extraction process effectively takes place in the region between the ergosphere boundary and the photon sphere radius.
The Keplerian angular velocity for particles is given as $\Omega=d\phi/d\tau / dt/d\tau$ \cite{cai2024analysis}
with
\begin{equation}
\Omega = \frac{-\partial_r g_{t\phi} \pm \sqrt{(\partial_r
g_{t\phi})^2 - (\partial_r g_{tt})(\partial_r
g_{\phi\phi})}}{\partial_r g_{\phi\phi}}.\label{keplarianvelocity}
\end{equation}
The positive sign corresponds to prograde orbits, while the negative sign corresponds to retrograde orbits. Therefore, we consider the Keplerian angular velocity for prograde motion and select the corresponding expression with the positive sign. Equation (\ref{keplarianvelocity}) characterized the circular orbits, i.e, it only involves the angular velocity in azimuthal direction, not in the radial direction.

\section{EXTRACTING RASN BLACK HOLE ENERGY IN THE CIRCULAR ORBIT REGION}

\subsection{Magnetic Reconnection Process in Circular Orbits}
\begin{figure}[H] \centering
\includegraphics[width=8cm,height=5cm]{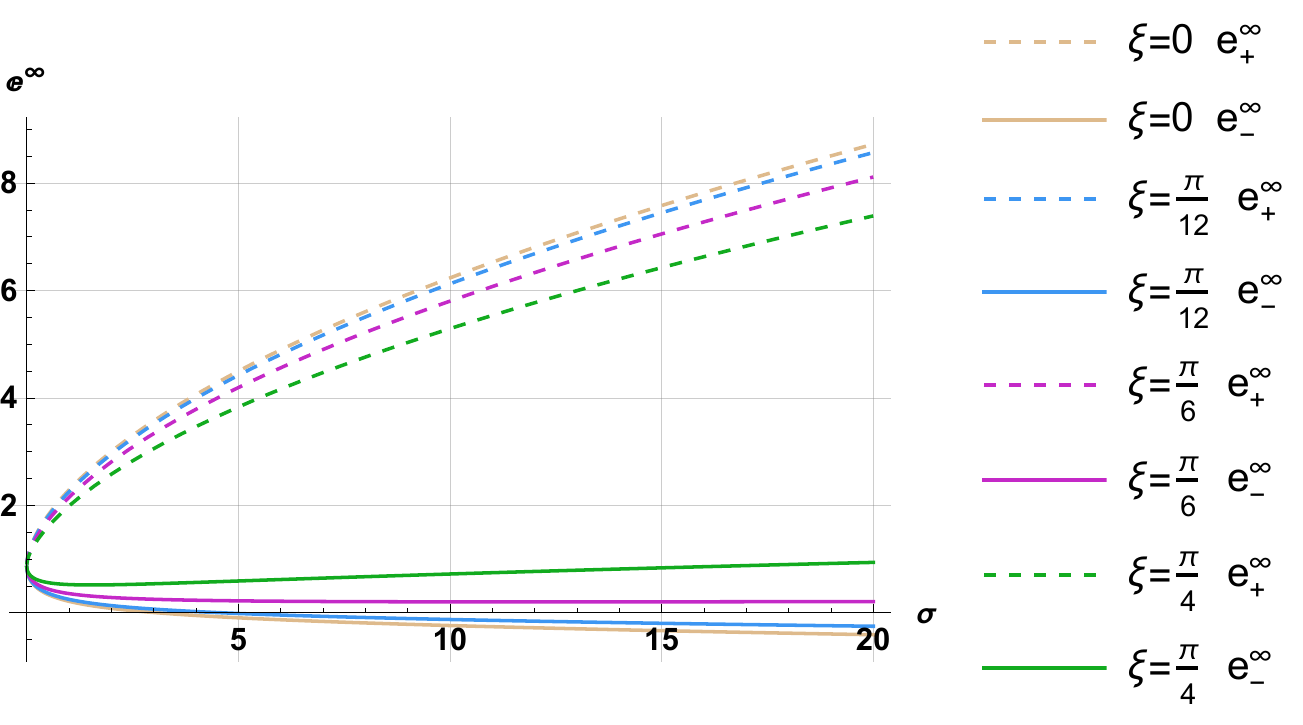}
\caption{Plot showing the variation of $e^{\infty}_{+}$ and $e^{\infty}_{-}$ with respect to $\sigma$ for different values of $\xi$, with fixed $a=0.98$, $K=1/100$,~$r=1.8$,~$N=4$, and $l=30$.} \label{fig2}
\end{figure}

\begin{figure}[H]
\begin{center}
\subfigure[~$K=1/100,~N=4,~a=0.98,~r=1.7,\xi=\pi/12$]{\includegraphics[width=8cm,height=5cm]{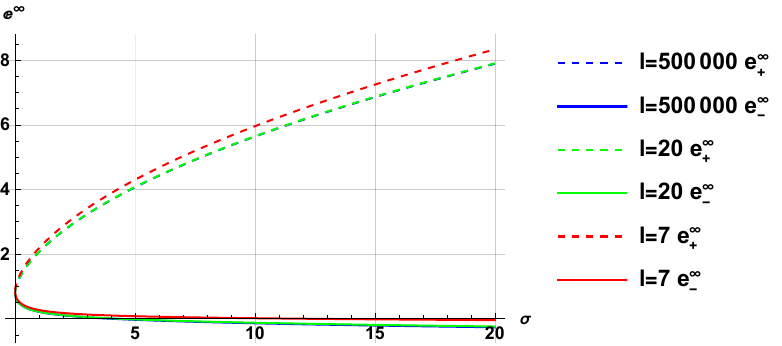}}
\subfigure[~$K=1/100,~N=4,~a=0.98,~r=2,~\xi=\pi/12$]{\includegraphics[width=8cm,height=5cm]{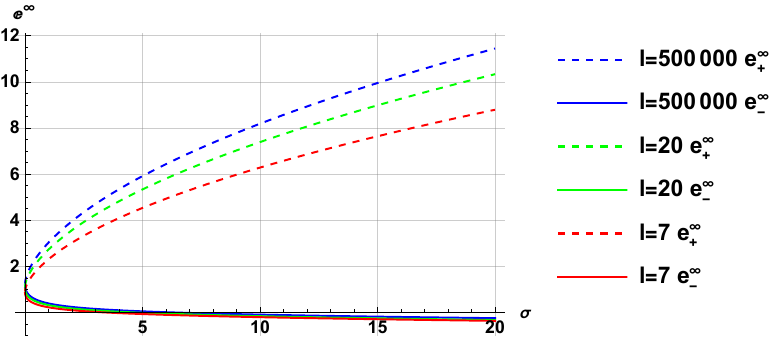}}
    \caption{Plot showing the variation of $e^{\infty}_{+}$ and $e^{\infty}_{-}$ with respect to $\sigma$ for different values of $l$.}
    \label{fig3}
    \end{center}
\end{figure}

\begin{figure}[H]
\begin{center}
\subfigure[~$K=1/100,~a=0.98,~l=20,~r=1.5,~\xi=\pi/12$]{\includegraphics[width=8cm,height=5cm]{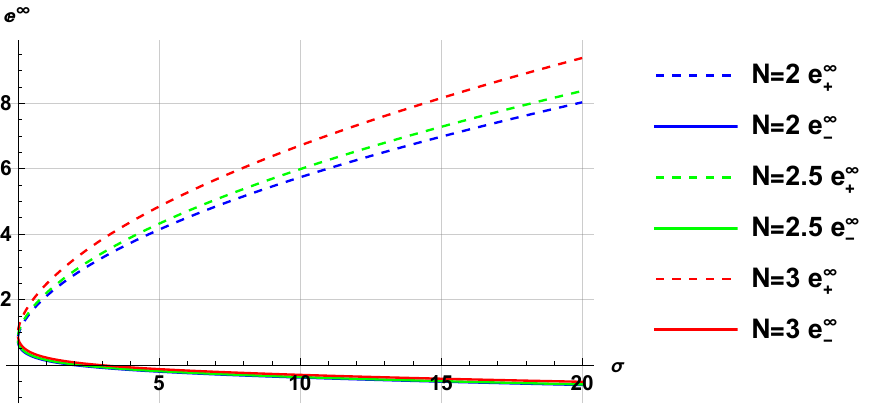}}
\subfigure[$N=2,~a=0.98,~l=7,~r=1.2,~\xi=\pi/12$]{\includegraphics[width=8cm,height=5cm]{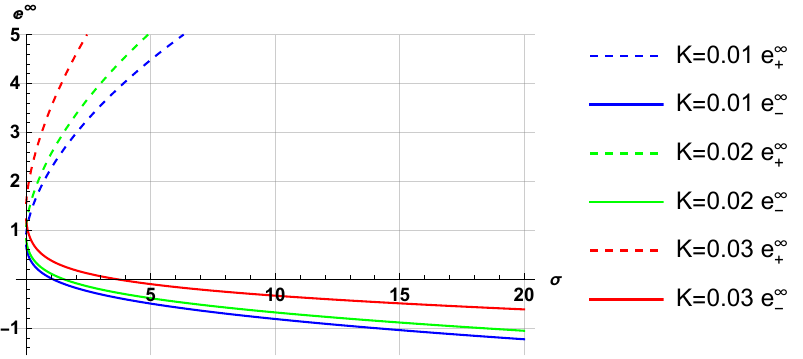}}
\caption{Plots showing the variation of $e^{\infty}_{+}$ and $e^{\infty}_{-}$ with respect to $\sigma$ for different values of $N$ (left panel), and for different values of $K$ (right panel).}\label{fig4and5}
    \end{center}
\end{figure}

To begin with, let us redefine some fundamental concepts related to magnetic reconnection in a stationary, axisymmetric spacetime. For the analysis, we adopt the Zero Angular Momentum Observer (ZAMO) frame \cite{bardeen1972rotating}. In this frame, the metric can be written as follows:
\begin{equation}
ds^2 = - d\hat{t}^{\,2} + \sum_{i=1}^{3} \left( d\hat{x}^i \right)^2
= \eta_{\mu \nu} d\hat{x}^\mu d\hat{x}^\nu,
\label{zamo}
\end{equation}
where
\begin{equation}
d\hat{t} = \alpha \, dt, \qquad
d\hat{x}^i = \sqrt{g_{ii}} \, dx^i - \alpha \beta^i dt ,
\label{zamocomponents}
\end{equation}
with the following quantities
\begin{equation}
\alpha = \left( -g_{tt} + \frac{g_{t\phi}^2}{g_{\phi\phi}} \right)^{1/2}, \qquad \beta^\phi = \frac{\sqrt{g_{\phi\phi}} \, \omega^\phi}{\alpha},\quad \beta^r = \beta^\theta = 0, \qquad
\omega^\phi =-\frac{g_{t\phi}}{g_{\phi\phi}}. \label{zamoquantaties}
\end{equation}
Thus the Keplerian velocity in the ZAMO frame is of the form
\begin{equation}
\hat{v}_K = \frac{1}{\alpha}
\left( \sqrt{g_{\phi\phi}} \, \Omega - \alpha \beta^\phi \right) .
\label{keplerianinZAMO}
\end{equation}
Within the energy-momentum tensor, we consider the single-fluid plasma approximation as
\begin{equation}
T^{\mu \nu} = p g^{\mu \nu} + w u^\mu u^\nu
+ F^{\mu \sigma} F^{\nu}_{\ \sigma}
- \frac{1}{4} g^{\mu \nu} F^{\alpha \beta} F_{\alpha \beta},
\label{momenttensor}
\end{equation}
where  $F$, $u$, $w$, $p$ are the electromagnetic field tensor, four velocity, plasma enthalpy density and pressure, respectively. We know the magnetic reconnection process is efficient, which means that the magnetic energy is completely converted into kinetic energy; because of this we can neglect the electromagnetic field tensor part.
So, the energy at infinity per enthalpy for the accelerated and decelerated plasma by using the adiabatic and incompressible nature of plasma is approximately given as \cite{comisso2021magnetic}
\begin{equation}
\begin{aligned}
e_{\pm}^{\infty} &= \frac{-\alpha g_{\mu 0} T^{\mu 0}}{w} \\ &= \alpha \hat{\gamma}_\kappa \left[
(1 + \beta^{\phi} \hat{v}_\kappa)(1+\sigma)^{1/2} \pm \cos \xi (\beta^{\phi} + \hat{v}_\kappa) \sigma^{1/2} - \frac{1(1+\sigma)^{1/2} \mp \cos \xi \hat{v}_\kappa \sigma^{1/2}} {4 \hat{\gamma}_\kappa^2 (1+\sigma - \cos^2 \xi \hat{v}_\kappa^2)} \right],
\end{aligned}
\label{energyatenthalpy}
\end{equation}
where $\hat{\gamma}_\kappa$ is the lorentz factor of $\hat{v}^2_\kappa$, which is in the form of
\begin{equation}
   \hat{\gamma}_\kappa= \frac{1}{\sqrt{(1-\hat{v}^2_\kappa)}},
   \label{lorentz}
\end{equation}
$\sigma$ is the plasma magnetization parameter and in the ,local rest frame $\xi$ is the azimuthal angle of a fluid. The relationship between enthalpy density and pressure of a relativistically hot plasma is $p=w/4$. For energy extraction, the following two conditions must satisfying
\begin{equation}
e_{-}^{\infty} < 0, \quad
\Delta e_{+}^{\infty} = e_{+}^{\infty} - \left[ 1 - \frac{\Gamma}{4(\Gamma - 1)} \right] = e_{+}^{\infty} > 0, \label{energyextractioncondition}
\end{equation}
which is similar to the Penrose process \cite{penrose1969gravitational}, and $\Gamma$  is the polytropic index, which is equal to $4/3$. In Figs. \ref{fig2}-\ref{fig4and5}, we interpret the $e^{\infty}_{+}$ and $e^{\infty}_{-}$, where $r$ is the dominant reconnection point, so-called the X-point \cite{comisso2021magnetic}.
In Fig. \ref{fig2}, we observe that $e^{\infty}_{+}$ always remains greater than $0$, whereas $e^{\infty}_{-}$ is generally less than $0$. It can be seen that, with increasing $\sigma$, the value of $e^{\infty}_{+}$ increases. In contrast, as the azimuthal angle $\xi$ increases, $e^{\infty}_{+}$ decreases. Similarly, $e^{\infty}_{-}$ also increases with increasing $\sigma$, while it decreases as $\xi$ increases.

In Fig. \ref{fig3}, we observe that, for different values of $l$, neither $e^{\infty}_{+}$ nor $e^{\infty}_{-}$ varies monotonically. Instead, the behavior of both $e^{\infty}_{+}$ and $e^{\infty}_{-}$ depends strongly on the location of the X-point. Moreover, for larger values of $r$, the differences between the dashed lines corresponding to different values of $l$ become more pronounced and clearly distinguishable.
The left panel of Fig. \ref{fig4and5} interpret that the increasing values of $N$ leads to an increase in $e^{\infty}_{+}$, while $e^{\infty}_{-}$ remains nearly unchanged for all values of $N$. Whereas from the right panel of Fig. \ref{fig4and5}, it can be seen that as $K$ increases, $e^{\infty}_{-}$ exhibits a slight decrease, although the variation is relatively small. This behavior is favorable for energy extraction. Additionally, $e^{\infty}_{+}$ is found to increase with increasing $K$.

\subsection{Parameter Space for Energy Extraction Via Magnetic Reconnection in Circular Orbits}

In Figs.{ \ref{fig6}-\ref{fig10}, we interpret the allowed energy region for energy extraction ($e^{\infty}_{-}<0$) in $r$ verses $a$ plane. In all panels, the purple dashed line represents the photon sphere radius, the blue solid line represents the ergosphere, the red solid line represents the event horizon, and from left to right, $\sigma$ corresponds to $100, 30, 10$ and $3$, respectively.
\begin{figure}[H]
\begin{center}
\subfigure[~$N=3$]{\includegraphics[width=4.7cm,height=4.3cm]{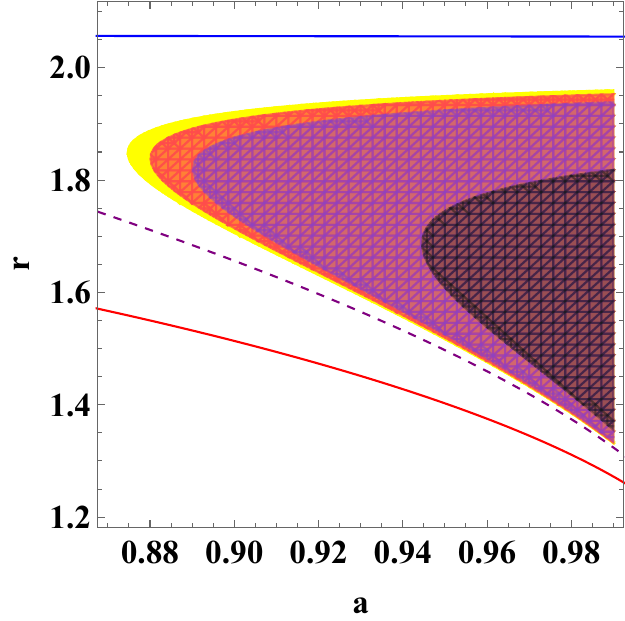}}
\subfigure[~$N=2$]{\includegraphics[width=4.7cm,height=4.3cm]{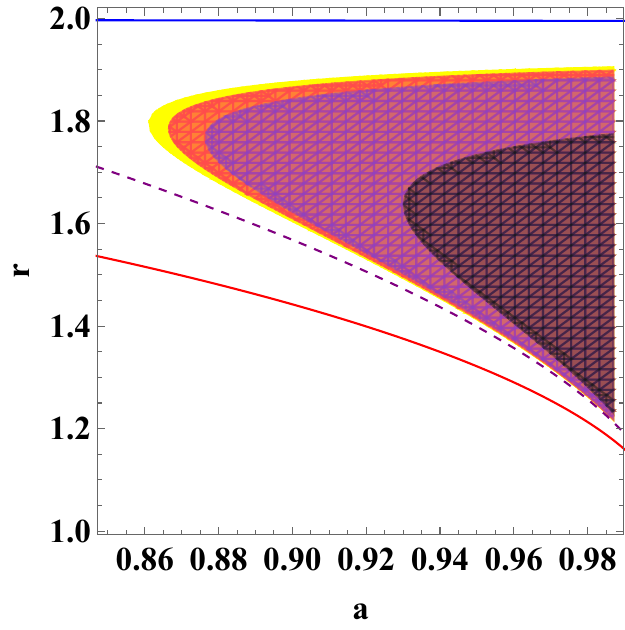}}
\subfigure[~$N=1$]{\includegraphics[width=4.7cm,height=4.3cm]{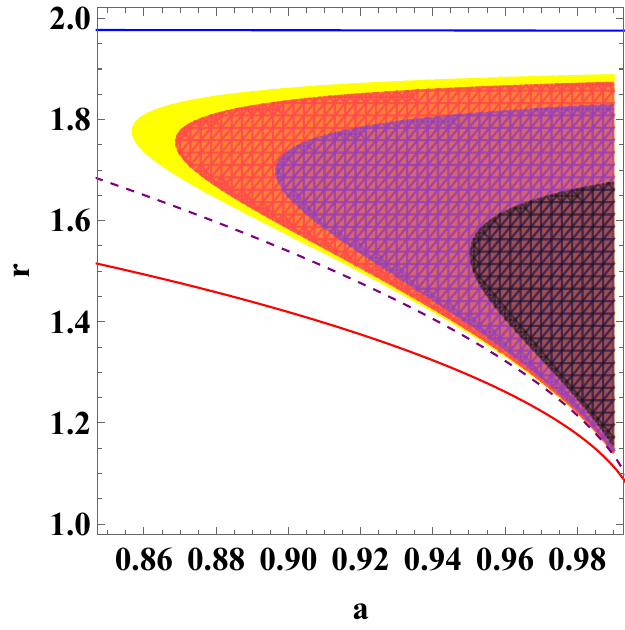}}
    \caption{Plots showing the allowed energy extraction regions for different values of $N$ with fixed $l=20,~K=1/100$ and $\xi=\pi/12$.}
    \label{fig6}
    \end{center}
\end{figure}

In Fig. \ref{fig6}, it can be observed that the allowed region for energy extraction becomes larger as $\sigma$ increases. These results are consistent with \cite{comisso2021magnetic}. Furthermore, decreasing the value of $N$ reduces the minimum allowed spin parameter, changing from $0.87$ in Fig. \ref{fig6} (a) to $0.85$ in Fig. \ref{fig6} (c). In contrast, the maximum allowed spin remains unchanged for all values of $N$. In addition, as the event horizon, photon sphere radius, and ergosphere expand, the location of the reconnection layer shifts outward accordingly.

\begin{figure}[H]
\begin{center}
\subfigure[~$l=10$]{\includegraphics[width=4.7cm,height=4.3cm]{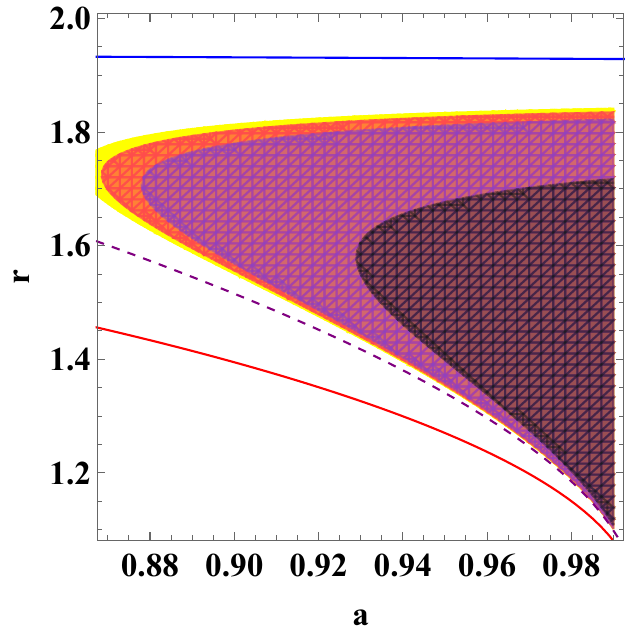}}
\subfigure[~$l=15$]{\includegraphics[width=4.7cm,height=4.3cm]{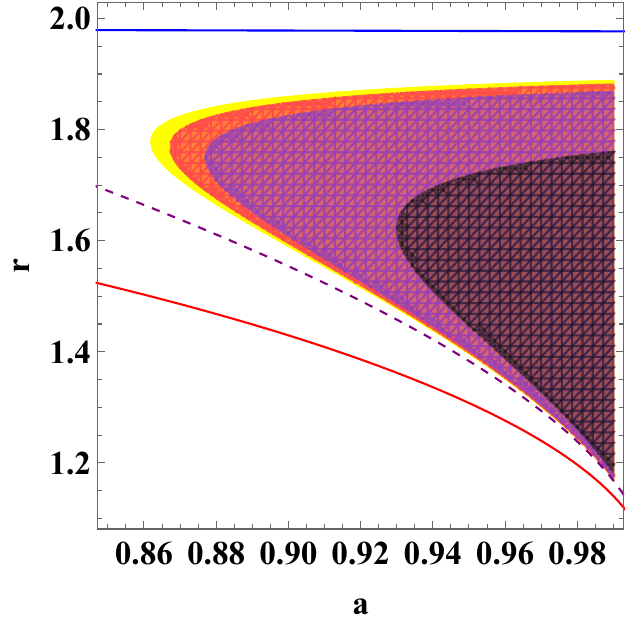}}
\subfigure[~$l=20$]{\includegraphics[width=4.7cm,height=4.3cm]{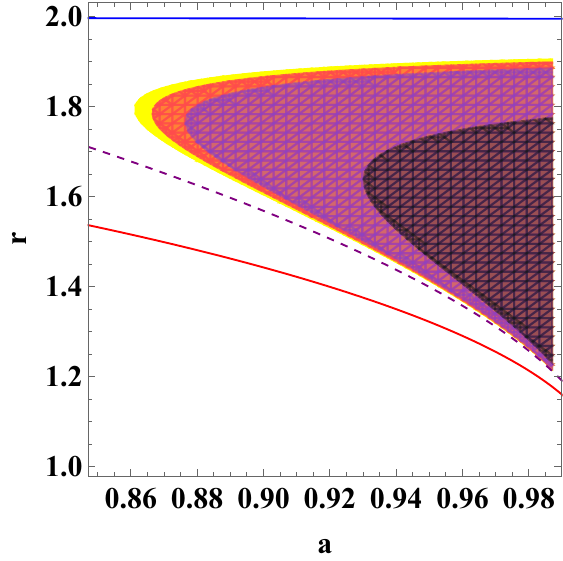}}
    \caption{Plots showing the allowed energy extraction regions for different values of $l$ with fixed $N=2,~K=1/100$ and $\xi=\pi/12$.}
    \label{fig7}
    \end{center}
\end{figure}
\begin{figure}[H]
\begin{center}
\subfigure[~$K=1/20$]{\includegraphics[width=4.7cm,height=4.3cm]{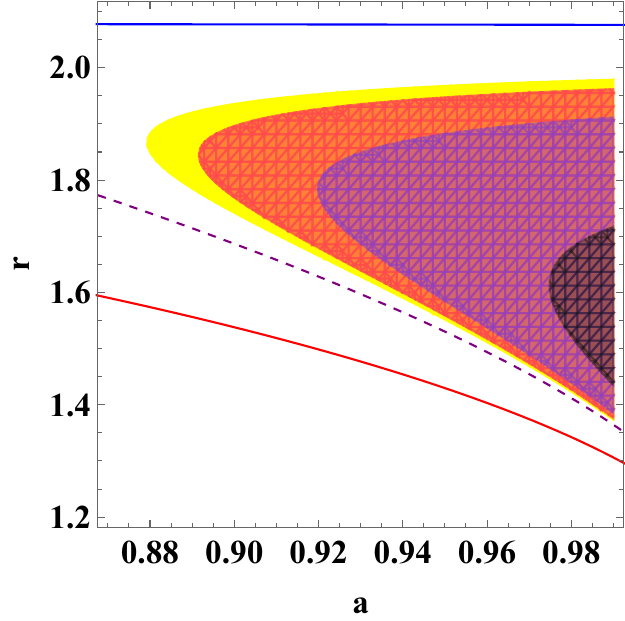}}
\subfigure[~$K=1/50$]{\includegraphics[width=4.7cm,height=4.3cm]{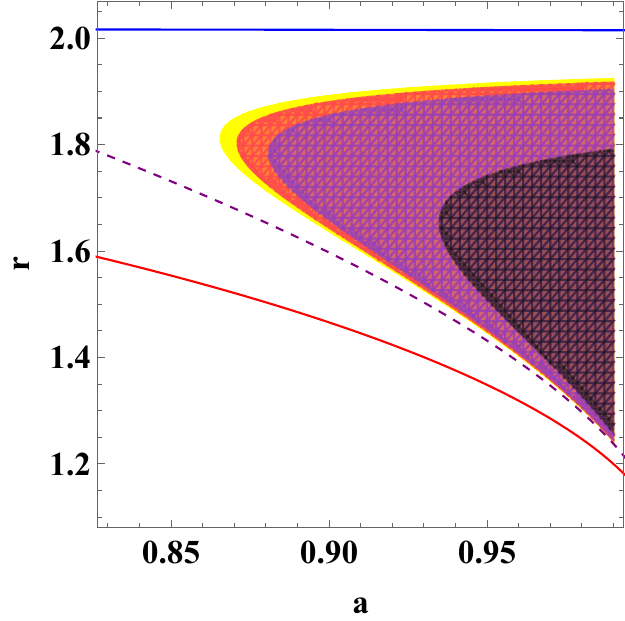}}
\subfigure[~$K=1/100$]{\includegraphics[width=4.7cm,height=4.3cm]{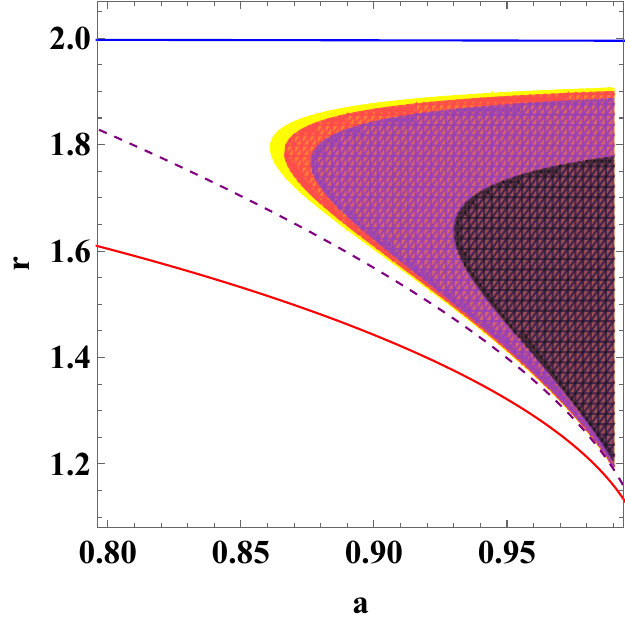}}
    \caption{Plots showing the allowed energy extraction regions for different
values of $K$ with fixed $\xi=\pi/12,~l=20$ and $N=2$.}\label{fig8newk}
    \end{center}
\end{figure}

In Fig. \ref{fig7}, it is evident that increasing the parameter $l$ lowers the minimum allowed spin, thereby permitting energy extraction even for smaller spin values. However, the maximum
allowed spin remains unchanged. Figure \ref{fig8newk} shows that decreasing $K$ also reduces the minimum allowed spin, from $0.86$ in Fig. \ref{fig8newk} (a) to $0.80$ in Fig. \ref{fig8newk} (c), while the maximum allowed spin stays constant. Moreover, with the increase in the size of the ergosphere, event horizon, and photon sphere radius, the reconnection layer shifts outward accordingly. Therefore, in the case of circular orbits, the parameters $N$, $l$, and $K$ all contribute to lowering the critical spin required for energy extraction.
\begin{figure}[H]
\begin{center}
\subfigure[~$\xi=\pi/12$]{\includegraphics[width=4.7cm,height=4.3cm]{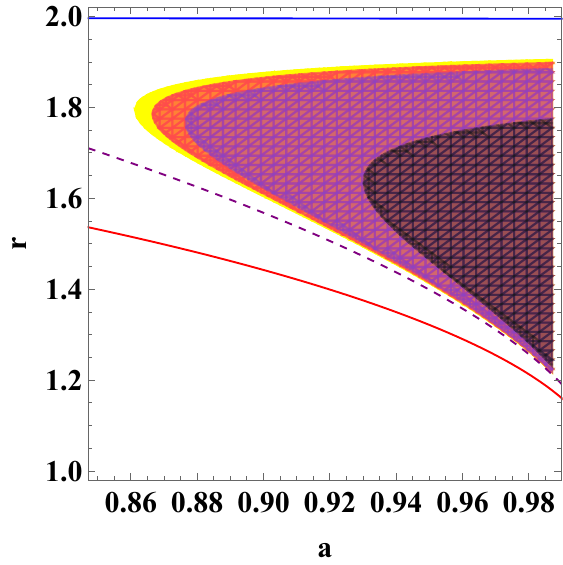}}
\subfigure[~$\xi=\pi/6$]{\includegraphics[width=4.7cm,height=4.3cm]{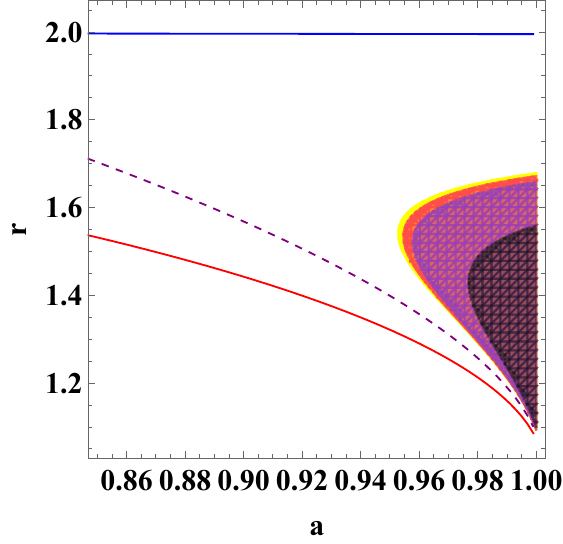}}
\subfigure[~$\xi=0$]{\includegraphics[width=4.7cm,height=4.3cm]{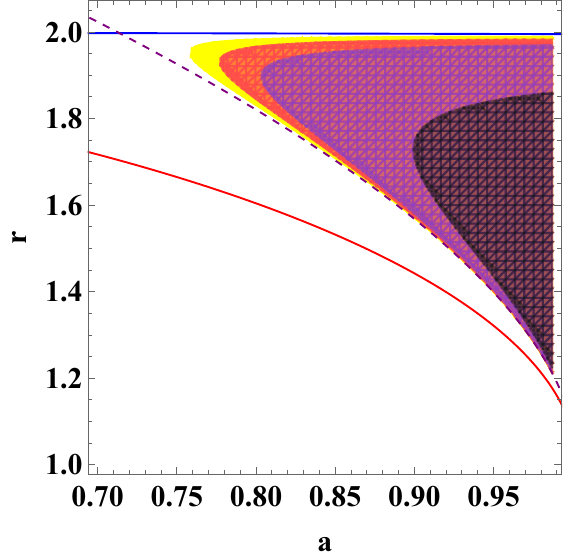}}
\caption{Plots showing the allowed energy extraction regions for
different values of $\xi$ with fixed $K=1/100,~l=20$ and
$N=2$.}\label{fig9}
\end{center}
\end{figure}

In Fig.~\ref{fig9}, we observe that as $\xi$ decreases, the area of the allowed energy extraction region increases, the minimum allowed spin decreases, and the position of the reconnection layer more pronounced. However, the overall behavior of the magnetic reconnection process remains unaffected by variations in $\xi$. Therefore, for conveniently we set $\xi=\pi/12$ in the subsequent analysis. In Fig.~{\bf \ref{fig10}}, we discuss the minimum value
of the spin parameter $a$ required for magnetic reconnection to occur. Our analysis shows that smaller values of $K$, $l$, and $N$ lead to lower critical values of the spin parameter. The results further indicate that the allowed spin range for energy extraction lies approximately between $0.70$ and $0.98$, demonstrating that the energy extraction process can take place even for relatively low black hole spins.
\begin{figure}[H]\centering
\includegraphics[width=6cm,height=5.6cm]{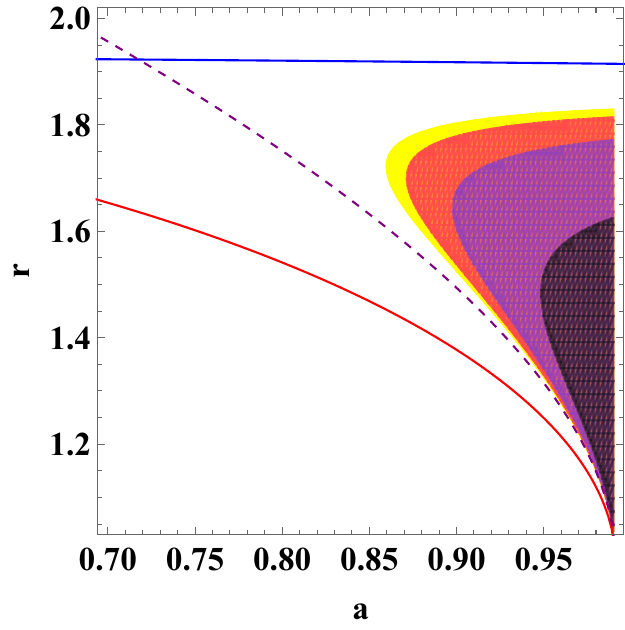}
\caption{Plot showing the allowed energy extraction region with fixed $l=10,~K=1/500,~N=2$ and $\xi=\pi/12$.} \label{fig10}
\end{figure}

\subsection{Power and Efficiency of Energy Extraction in Circular Orbits}

After investigating the feasibility of energy extraction, we now compare the corresponding power and efficiency of the process. The power related with energy extraction is given by
\cite{comisso2021magnetic}.
\begin{equation}
P=-e^{\infty}_{-}wA_{in}U_{in}\label{power}
\end{equation}
Here $U_{in}$ is used  for two conditions, such as for collisionless condition  $U_{in} \approx0.1$ \cite{comisso2016value}, and for collisional condition $U_{in} \approx0.01$
\cite{huang2010scaling,comisso2016visco,uzdensky2010fast}. In this paper, we use
$U_{in}=0.1$. $A_{in} $ indicates the cross-sectional area of the inflowing plasma, which is defined as
\begin{equation}
A_{in} \sim (r^2_{E}-r^2_{ph}), \label{crossArea}
\end{equation}
in which $r_{ph}$ represent the radius of photon sphere and $r_{E }$ indicates the ergosphere boundary. In Figs.~\ref{fig11} and \ref{fig12}, we present the energy extraction power per enthalpy density, $P/w$ with respect to $r$. The red, orange, blue, and green solid curves correspond to $\sigma=3$, $10$, $30$, and $100$, respectively. The green dashed, purple dashed, and red dashed lines denote the boundaries of the ergosphere, photon sphere, and event horizon, respectively.

\begin{figure}[H]
\begin{center}
\subfigure[~$N=2,~K=1/100,~l=20,~\xi=\pi/12,~a=0.98$]{\includegraphics[width=8cm,height=5.5cm]{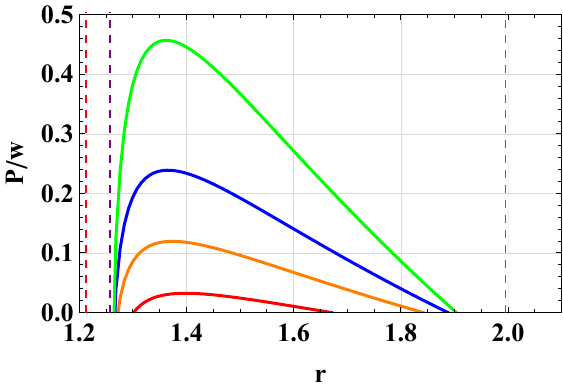}}
\subfigure[~$N=3,~K=1/75,~l=25,~\xi=\pi/12,~a=0.97$]{\includegraphics[width=8cm,height=5.5cm]{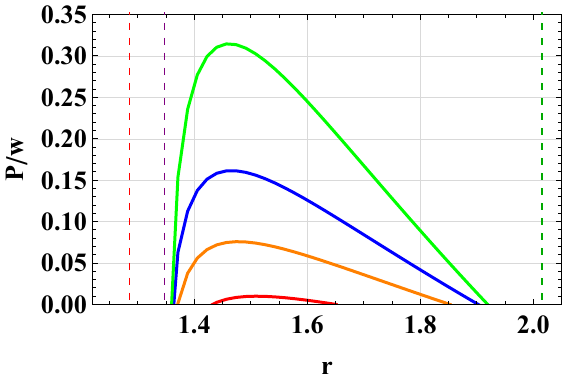}}
\caption{Plots showing the energy extraction power for different
values of $\sigma$.}\label{fig11}
\end{center}
\end{figure}
From Fig.~\ref{fig11} (a), we found that the energy extraction power starts from the outside of the photon sphere radius, which is consistent with the characteristics of circular orbits. The power initially increases with $r$, reaches a maximum value, and then gradually decreases. Moreover, the power increases with the aid of $\sigma$. Upon comparing Figs.~\ref{fig11} (a) and \ref{fig11} (b), it can be observe that, for the same value of $\sigma$, the energy extraction power decreases as $N$ increases. It should also be seen that the spin parameter $a$ cannot be chosen identically in both cases because the allowed spin range of the black hole depends on the value of $N$. A similar behavior is observed in Ref.~\cite{li2023energy}, where the allowed spin range also changes for different values of $h_0$. Comparing Figs.~\ref{fig11} (a) and \ref{fig12} (a), we found that the energy extraction power decreases as $l$ decreases, mainly due to the corresponding reduction in the allowed black hole spin. Again, the spin parameter
$a$ cannot remain the same because its allowed range varies with $l$. Finally, Fig.~\ref{fig12} (b) shows that the energy extraction power becomes significantly smaller for specific  combinations of the parameters $K$, $l$, and $N$.

\begin{figure}[H]
\begin{center}
\subfigure[~$N=2,~K=1/100,~l=7,~\xi=\pi/12,~a=0.95$]{\includegraphics[width=8cm,height=5.5cm]{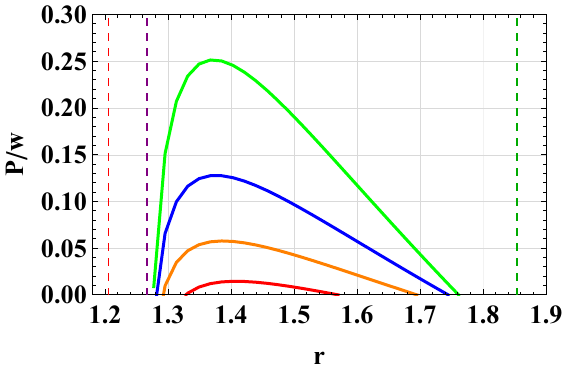}}
\subfigure[~$N=4,~K=1/100,~l=20,~\xi=\pi/12,~a=0.96$]{\includegraphics[width=8cm,height=5.5cm]{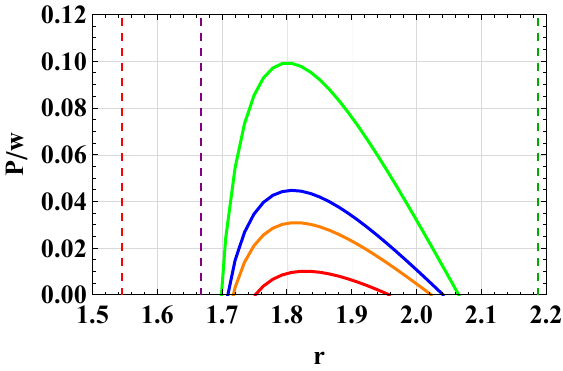}}
\caption{Plots showing the energy extraction power for different
values of $\sigma$.}\label{fig12}
\end{center}
\end{figure}
Now, we investigate the energy extraction efficiency, which is defined as follows
\begin{equation}
  \eta=\frac{e^\infty_{+}}{e^\infty_{+}+e^\infty_{-}}.
  \label{neta efficiency}
\end{equation}
To calculate the energy extraction, the conditions  $e^\infty_{+}>0$ and $e^\infty_{-}<0$ must be holds and the efficiency is always greater than $1$. In Figs. \ref{fig13} and \ref{fig14}, we
interpret the energy extraction efficiency with respect to $r$, while fixing $\sigma=100$ and $\xi=\pi/12$.
\begin{figure}[H]
\begin{center}
\subfigure[~$N=2,~K=1/150,~l=10$]{\includegraphics[width=8cm,height=5cm]{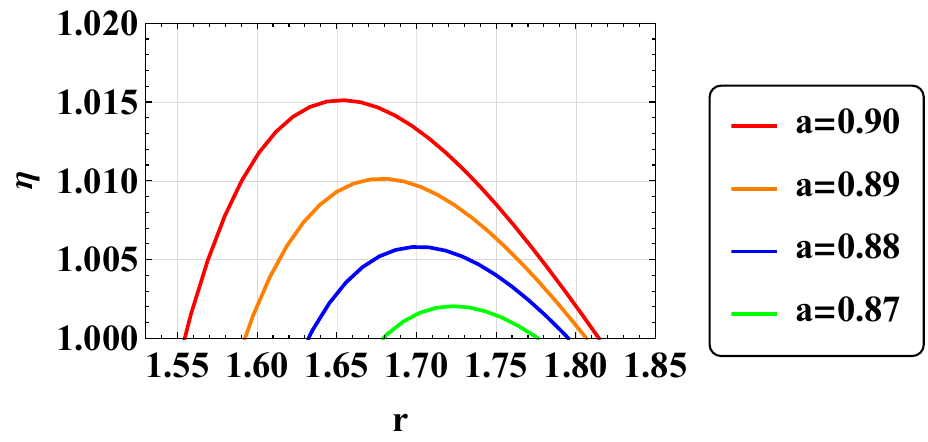}}
\subfigure[~$N=2,~K=1/90,~l=10$]{\includegraphics[width=8cm,height=5cm]{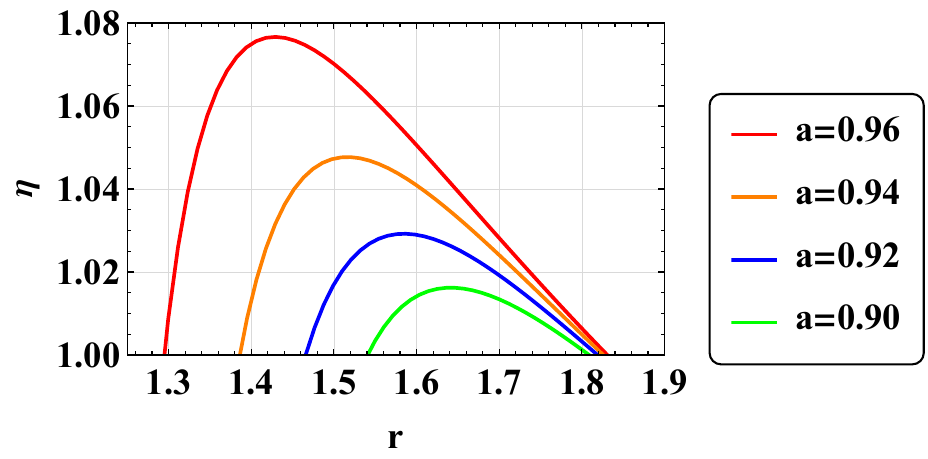}}
    \caption{Plots showing the energy extraction efficiency for different values of $a$.}
   \label{fig13}
    \end{center}
\end{figure}

\begin{figure}[H]
\begin{center}
\subfigure[~$N=3,~K=1/100,~l=5$]{\includegraphics[width=8cm,height=5cm]{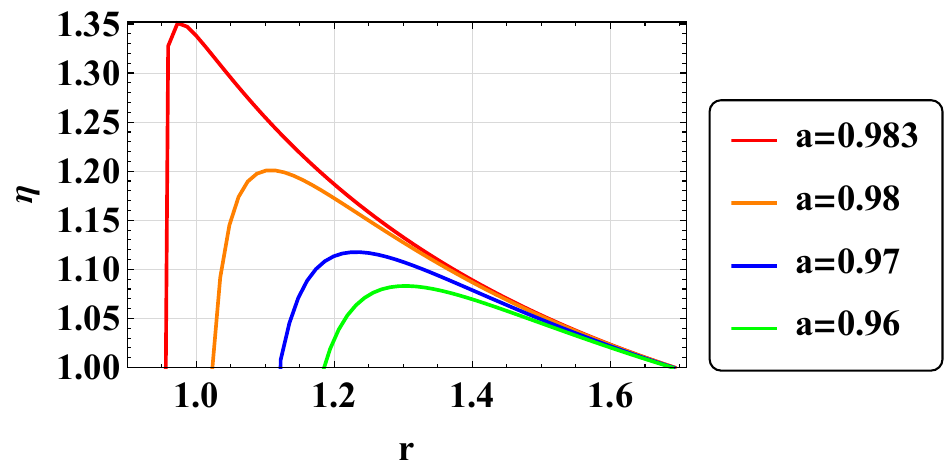}}
\subfigure[~$N=4,~K=1/130,~l=5$]{\includegraphics[width=8cm,height=5cm]{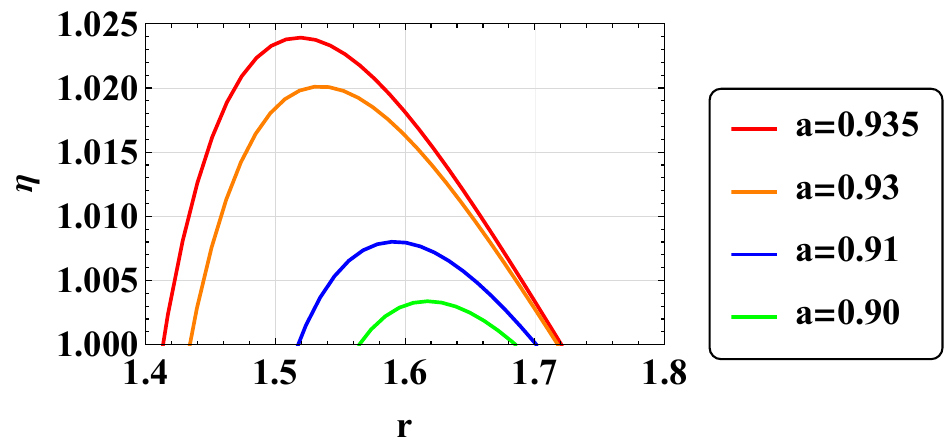}}
\caption{Plots showing the energy extraction efficiency for
different values of $a$.} \label{fig14}
\end{center}
\end{figure}

In Fig. \ref{fig13}(a), we illustrate that the energy extraction efficiency for different values of $a$ with $N=2,~K=1/150$ and $l=10$. The efficiency initially increases with $r$, reaches the
peak position, and then gradually decreases. It is also observed that the larger values of $a$ lead to higher energy extraction efficiency. Figure \ref{fig13}(b) present that the efficiency for different values of $a$ with $N=2,~K=1/90$ and $l=10$. From both panels of Fig. \ref{fig13}(a) and (b), we observe that increasing $K$ enhances the energy extraction efficiency due to the
corresponding increase in the allowed black hole spin, which is similar to the behavior observed for the power.

It should be noted that the values of a used in Figs. \ref{fig13}(a) and \ref{fig13}(b) are not same because the allowed spin range of the black hole changes for different values of $K$. We can also discuss that the influence of $l$ on the energy extraction efficiency. From the comparison between Fig. \ref{fig13}(a) and Fig. \ref{fig14}(a), it can be notice that decreasing $l$ reduces the energy extraction efficiency as a consequence of the lower spin, following a trend similar to that of the power. Moreover, in Fig. \ref{fig14}(b), we interpret the efficiency for several specific values of $l,~K$ and $N$. Comparing Fig. \ref{fig13}(a) with Fig. \ref{fig14}(b), we observe that the efficiency becomes higher, which again follows the same overall trend as observed in the power. Now we compare the energy extraction power ratio with that of the Blandford-Znajek framework \cite{blandford1977electromagnetic}. The corresponding Blandford-Znajek energy extraction power can be defined as \cite{tchekhovskoy2010black,camilloni2022blandford}:

\begin{equation}
P_{\text{BZ}} = \frac{\kappa_0}{16\pi} \, \Phi_H^2 \big(\Omega_H^2+\alpha_{1}\Omega_H^4
+\alpha_{2}\Omega_H^6\big),\label{blandfordpower}
\end{equation}
in which $\kappa_0=0.05$, $\alpha_1=1.38$ and $\alpha_2=-9.2$ are numerical constants, and $\Omega_H$ correspond to the angular velocity of the event horizon, which is expressed as
\begin{equation}
\Omega_H = \left. \frac{-g_{t\phi}}{g_{\phi\phi}} \right|_{r=r_+} = \frac{a\Sigma}{r_+^2  + a^2}. \label{eventhorizonvelocity}
\end{equation}
In Eq. (\ref{blandfordpower}) $\Phi_H$ represent the magnetic flux threading one hemisphere of the black hole event horizon, which has the following form as
\begin{equation}
\Phi_H = \frac{1}{2} \iint |B^r| \sqrt{g_{\theta\theta} g_{\phi\phi}} \, d\theta d\phi = \frac{2\pi \left( r_+^2  + a^2 \right)}{\Sigma} \, B_0 \sin\xi, \label{magneticflux}
\end{equation}
where $B_0={(w\sigma})^{1/2}$. Now, we calculate the power ratio by following a straightforward mechanism. Specifically, by squaring the above expression, yields
\begin{equation}
\Phi_H^2 = \frac{4\pi^2 \left( r_+^2  + a^2 \right)^2}{\Sigma^2} \, B_0^2 \sin^2\xi,
\label{squaringflux}
\end{equation}
Substituting $\Phi_H^2$ in Eq. (\ref{blandfordpower}), we obtain \begin{align}
P_{\text{BZ}} &= \frac{\kappa_0}{16\pi} \left( \frac{4\pi^2 \left( r_+^2  + a^2 \right)^2}{\Sigma^2} \right)  B_0^2 \sin^2\xi \left( \Omega_H^2+ \alpha_1 \Omega_H^4 + \alpha_2  \Omega_H^6 \right).
\label{Bzpowerafterfluxsubstitute}
\end{align}
After simplifying, we have
\begin{equation}
P_{\text{BZ}} = \frac{\kappa_0 \pi}{4} \frac{\left( r_+^2  + a^2 \right)^2}{\Sigma^2} \, B_0^2 \sin^2\xi \left( \Omega_H^2+ \alpha_1 \Omega_H^4 + \alpha_2 \Omega_H^6 \right). \label{constantputting}
\end{equation}

So, the reconnection power is given by $P = - e_0 A_{\text{in}} U_{\text{in}} \, \omega$. By dividing this expression by the power $P_{\text{BZ}}$, we obtain the corresponding normalized power ratio as follows
\begin{align}
\frac{P}{P_{\text{BZ}}} &= \frac{- e_0 A_{\text{in}} U_{\text{in}} \, \omega}{\frac{\kappa_0 \pi}{4} \frac{\left( r_+^2 + + a^2 \right)^2}{\Sigma^2} \, \omega \sigma \sin^2\xi \left(\Omega_H^2+
\alpha_1 \Omega_H^4 + \alpha_2 \Omega_H^6 \right)},\label{powerPBSpower}
\end{align}
which leads to
\begin{equation}
\frac{P}{P_{\text{BZ}}}=\frac{-4 e_0 A_{\text{in}} U_{\text{in}} \Sigma^2} {\kappa_0 \pi \sigma \left( \Omega_H^2+ \alpha_1 \Omega_H^4 + \alpha_2 \Omega_H^6 \right) \sin^2\xi \left( r_+^2+a^2
\right)^2}. \label{powerratio}
\end{equation}
The Eq. (\ref{powerratio}) shows that, for sufficiently small values of the azimuthal angle, the power ratio can exceed unity. This suggest that, under such conditions, the magnetic reconnection
mechanism can yield a higher energy extraction power than the Blandford-Znajek framework. So, for conveniently we fix $\xi=\pi/12$.

For instance, using the parameter set $l=20, K=1/100, N=2, \sigma=7, a=0.98 $ and $ r=1.4$, we obtain a power ratio of $5.143$. Since $5.143>1$, this clearly interprets that the energy extraction power from magnetic reconnection exceeds that of the Blandford-Znajek framework for this configuration. Notably, even in regimes characterized by relatively low spin and significant deviations, the power ratio remains greater than unity. For example, when $K=1/60, N=2, l=7, a=0.69, r=2$ and $\sigma=3$, the power ratio is $4.60587$. Although the absolute values of both power and efficiency decrease in this scenario, the magnetic reconnection mechanism still
dominates over the Blandford-Znajek mechanism. This trend can be attributed to the influence of parameters $l, K$, and $N$, which effectively lower the spin threshold and highlight the enhanced
efficiency of the magnetic reconnection mechanism.

\section{EXTRACTING RASN BLACK HOLE ENERGY IN THE PLUNGING REGION}

\subsection{Magnetic Reconnection Process in the Plunging Region}

Previously, we considered plasma moving on circular orbits lies beyond the position of the photon sphere. Closely followed by \cite{shen2024energy}, in this section, we assume that the plasma
initially moves on circular orbits beyond the innermost stable circular orbit (ISCO). Consequently, the plasma begins to plunge inward starting from the ISCO radius. At ISCO the circular orbits are unstable. The region $r<\hat{r_I}$, so-called the plunging region, where $\hat{r_I}$ indicates the ISCO radius, which is typically larger than the radius of photon sphere
\cite{wilkins2020venturing}. The Keplerian velocity formula given in Eq.~(\ref{keplarianvelocity}), which contains only the azimuthal component, is no longer feasible in the plunging region because the plasma also possesses a radial velocity component. In this case, we again impose the convenient ZAMO frame. The relationship between the BL frame and the four-velocity in the ZAMO frame is given by

\begin{equation}
\hat{U}^{\mu} = \hat{\gamma}_s \{1, \hat{v}_s^r, 0, \hat{v}_s^\phi\} = \left\{ \frac{E - \omega^\phi L}{\alpha}, \, \sqrt{g_{rr}} U^r, \, 0, \, \frac{L}{\sqrt{g_{\phi\phi}}} \right\}, \label{BL4vrelation}
\end{equation}
where
\begin{equation}
(U^r)^2=\left( \frac{dr}{d\tau} \right)^2 = \frac{R(r)}{\rho^4}, \label{Uofr}
\end{equation}
\begin{equation}
\rho^2 \left( \frac{dt}{d\tau} \right) = \frac{E(r^2 +  a^2)^2 - a(r^2 + a^2)L\Sigma}{\Delta_r}
- \frac{\sin^2\theta}{\Delta_\theta} \left( a^2 E - \frac{aL\Sigma}{\sin^2\theta} \right),
\label{rahofunction}
\end{equation}
\begin{equation}
E = - (g_{tt} + \Omega g_{t\phi}) \left( \frac{dt}{d\tau} \right),\quad L = (g_{t\phi} + \Omega g_{\phi\phi}) \left( \frac{dt}{d\tau} \right).\label{ener and mumt}
\end{equation}
At ISCO, the conserved quantities $E$ and $L$ are denoted as
\begin{equation}
E_I = E(r_I), \quad L_I = L(r_I). \label{ISCOLE}
\end{equation}
From Eq.~(\ref{Uofr}), the expression of $U^{r}=\pm \frac{\sqrt{R(r)}}{\rho^2}$, and at ISCO, it has the following expression as
\begin{equation}
U^r = -\frac{1}{\rho^2} \sqrt{ \Sigma^2[(r^{2}+a^{2})E_I - aL_I\big]^2 - \Delta_r\big[(L_I-aE_I)^2 +\mathcal{O} ]},
\label{Uafterradial}
\end{equation}
where $-$ve sign indicates the infalling motion. Substitute Eq. (\ref{Uafterradial}) into Eq. (\ref{BL4vrelation}), we obtain
\begin{equation}
\begin{aligned}
\hat{U}^{\mu} &=\hat{\gamma_s} \left\{ \frac{E - \omega^\phi L}{\alpha}, \, -\frac{\sqrt{g_{rr}}}{\rho^2} \sqrt{ -\Delta_r\big[(L_I - aE_I)^2 + \mathcal{O}\big] + \Sigma^2 \big[(r^{2}+a^{2})E_I - aL_I\big]^2 }, \right. \\ &\quad \left. 0, \, \frac{L}{\sqrt{g_{\phi\phi}}} \right\}
\label{BL4vafterradial}
\end{aligned}
\end{equation}
gives $\hat{v}_s^{(r)}$, $\hat{v}_s^{(\phi)}$, and $\hat{v_s}=\sqrt{( \hat{v}_s^{(r)})^2+ (\hat{v}_s^{(\phi)})^2},$ $\hat{\gamma_s}$ represent the Lorentz factor of $\hat{v_s}$, $r_I$
satisfies (\ref{conditionforR })  and also $ \frac{d^2R(r)}{dr^2}=0$. In this scenario, the $e^\infty_\pm$ of (\ref{energyatenthalpy}) modify as \cite{chen2024energy}
\begin{equation}
\begin{split}
e_{\pm}^{\infty} &= \alpha \hat{\gamma}_s \gamma_{\mathrm{out}} \bigg( (1 + \beta^{\phi} \hat{v}_s^{\phi}) \pm v_{\mathrm{out}} \left( \hat{v}_s + \beta^{\phi}\frac{\hat{v}_s^{(\phi)}}{\hat{v}_s} \right)\cos\xi \mp v_{\mathrm{out}} \beta^{\phi} \frac{\hat{v}_s^{(r)}}{\hat{\gamma}_s \hat{v}_s}\sin\xi \bigg) \\ &\quad -\frac{\alpha}{4\hat{\gamma}_s \hat{\gamma}_{\mathrm{out}}
\left(1 \pm \hat{v}_s \hat{v}_{\mathrm{out}} \cos\xi\right)},
\end{split}
\label{Epsilon}
\end{equation}
where $\gamma_{\mathrm{out}}$ is the Lorentz factor and ${v}_{\mathrm{out}} $ indicates the outflow speed, which has the following expressions as
\begin{equation}
{\gamma}_{\mathrm{out}}=\sqrt{\sigma+1},\quad {v}_{\mathrm{out}}=\sqrt{\frac{\sigma} 1+\sigma}. \label{Lorentzfactor}
\end{equation}
Now, we interpret the allowed energy extraction region in the $r-a$ plane for the plunging region in Figs.~\ref{fig20}-\ref{fig24}. In these figures, the black line corresponds to the ISCO, the purple line represents the radius of photon sphere, the blue line represents the ergosphere, and the red line represents the event horizon. From left to right, the values of $\sigma$ are $100, 30, 10$, and $3$, respectively.

\begin{figure}[H]
\begin{center}
\subfigure
[~$N=3$]{\includegraphics[width=4.7cm,height=4.3cm]{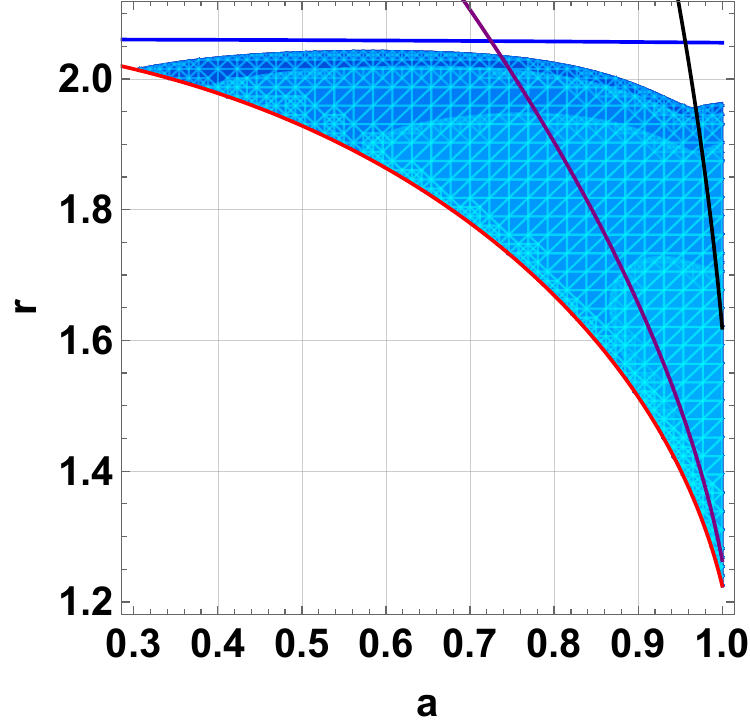}}
\subfigure[~$N=2$]{\includegraphics[width=4.7cm,height=4.3cm]{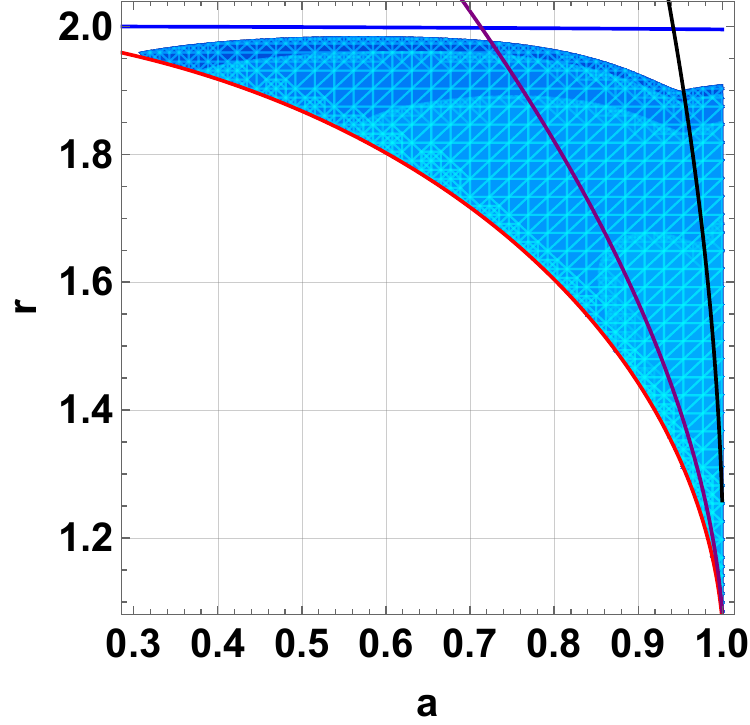}}
\subfigure[~$N=1$]{\includegraphics[width=4.7cm,height=4.3cm]{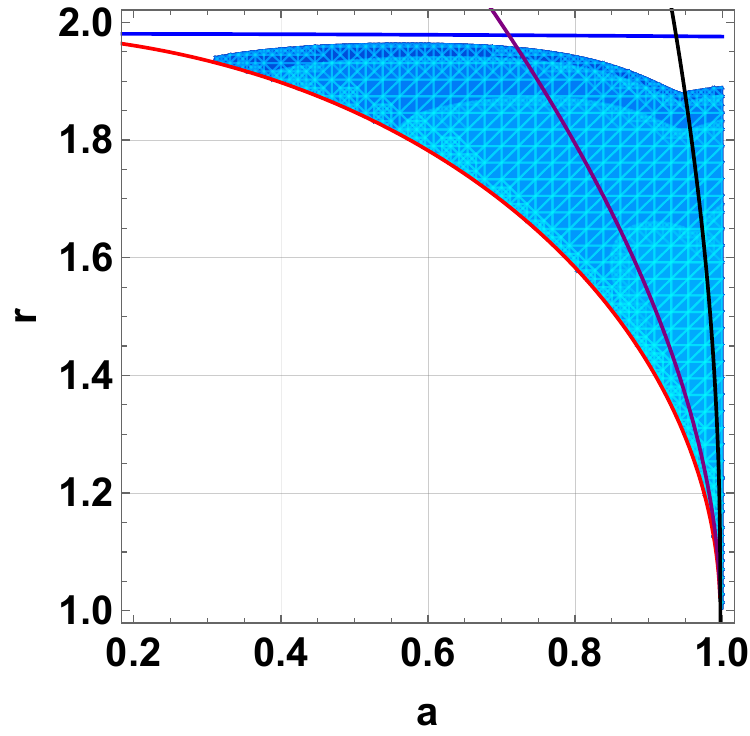}}
\caption{Plots showing the allowed energy extraction regions for
different values of $N$ with fixed $K=1/100,~\xi=\pi/12$ and $l=20$
in the plunging region.}\label{fig20}
\end{center}
\end{figure}

In Fig.~\ref{fig20}, we observe that as $\sigma$ decreases, the allowed energy extraction region also decreases. By comparing with Fig.~\ref{fig6}(a), we find that for $r>r_I$, the allowed energy extraction region coincides with that of the circular orbit case. However, for the condition $r<r_I$, the allowed energy extraction region in the plunging case is much greater than that in the circular orbit scenario. We also found that the minimum spin required for energy extraction in the plunging region is significantly lower than that in the circular orbit case, while the reconnection point is located at a higher values of radius $r$. For instance, when $\sigma=100$, the minimum spin allowing energy extraction in circular orbits (Fig.~\ref{fig6}(c)) is $0.85$, whereas in the plunging region the minimum spin decreases to $0.2$. Interestingly, from Fig.~\ref{fig20}, we notice that as $N$ decreases from left to right, the minimum allowed spin for energy extraction also decreases, changing from $0.3$ in Fig.~\ref{fig20}(a) to $0.2$ in Fig.~\ref{fig20}(c).

\begin{figure}[H]
\begin{center}
\subfigure[~$l=20$]{\includegraphics[width=4.7cm,height=4.3cm]{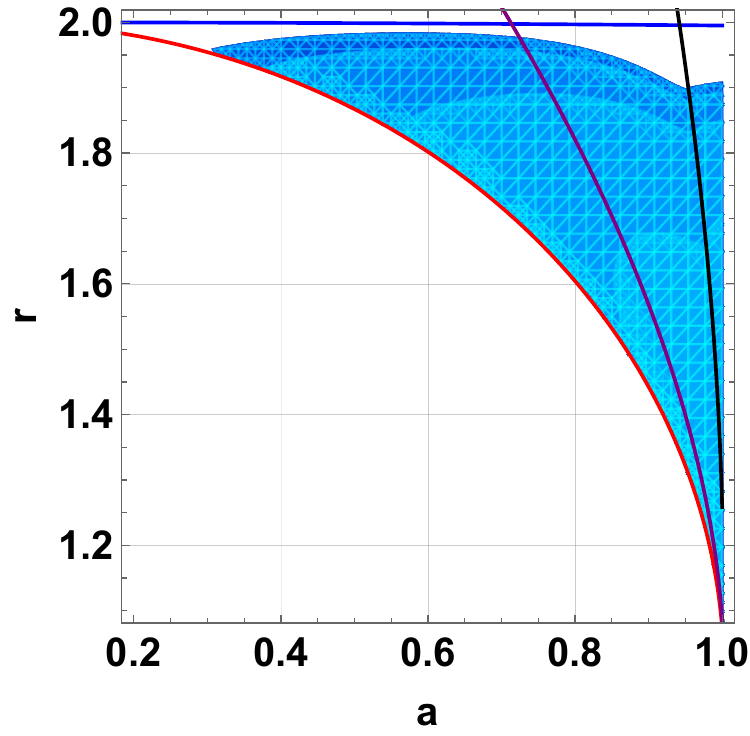}}
\subfigure[~$l=15$]{\includegraphics[width=4.7cm,height=4.3cm]{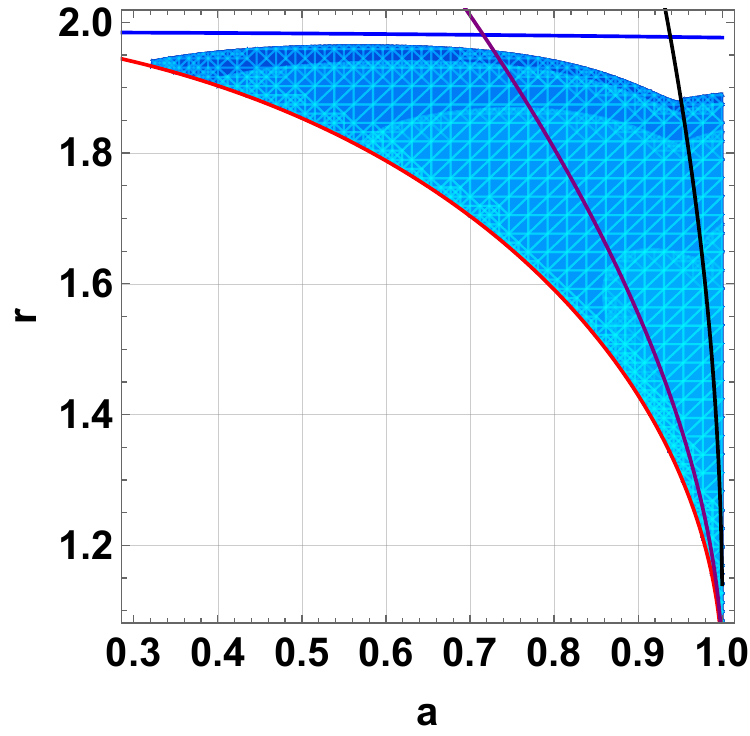}}
\subfigure[~$l=11$]{\includegraphics[width=4.7cm,height=4.3cm]{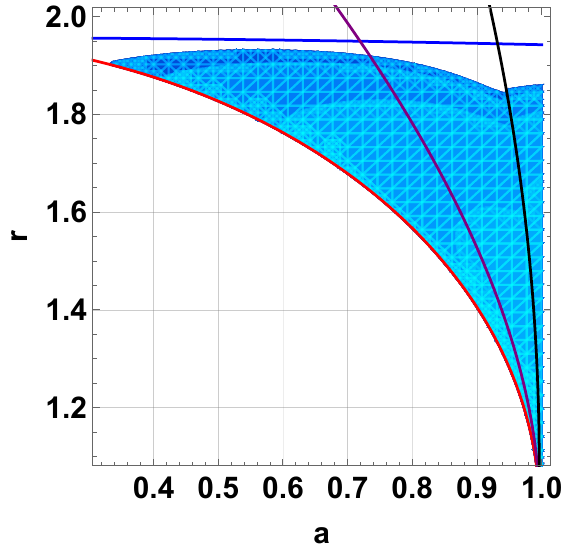}}
\caption{Plots showing the allowed energy extraction regions for
different values of $l$ with fixed $N=2,~\xi=\pi/12$ and $K=1/100$ in the plunging region.} \label{fig21}
\end{center}
\end{figure}

From Fig.~\ref{fig21}, we observe that as $l$ decreases, the corresponding minimum allowed spin for energy extraction slightly decreases, which is inconsistent with the trend observed in the
circular orbit case. Moreover, the minimum allowed spin for energy extraction in the plunging region is lower than that in the circular orbit scenario. For example, when $\sigma=100$ and $l=11$, the minimum allowed spin for energy extraction in the plunging region (Fig.~\ref{fig21}(c)) is $0.4$, whereas for circular orbits (Fig.~\ref{fig7}(c)) it is $0.85$. This demonstrates the advantage of the plunging region for energy extraction.

\begin{figure}[H]
\begin{center}
\subfigure[~$K=1/20
$]{\includegraphics[width=4.7cm,height=4.3cm]{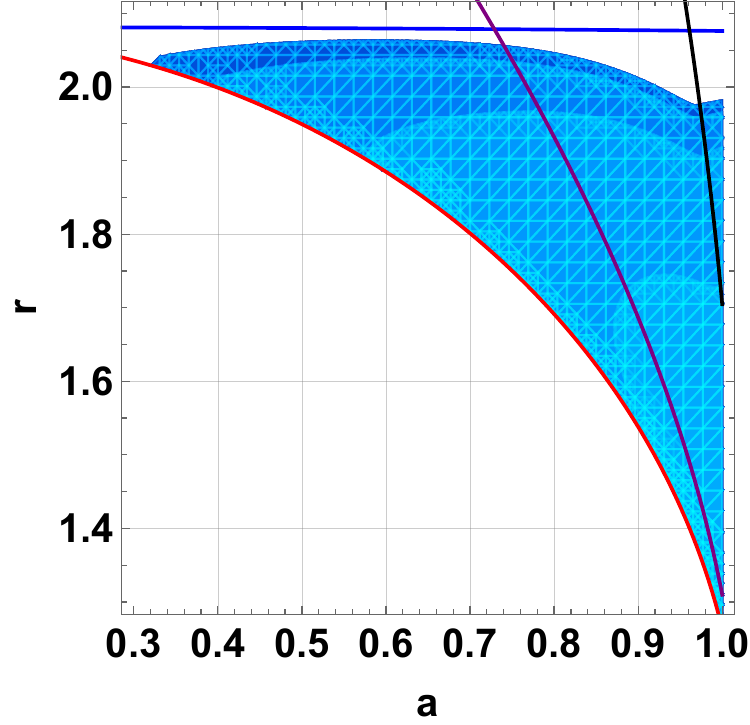}}
\subfigure[~$K=1/50
$]{\includegraphics[width=4.7cm,height=4.3cm]{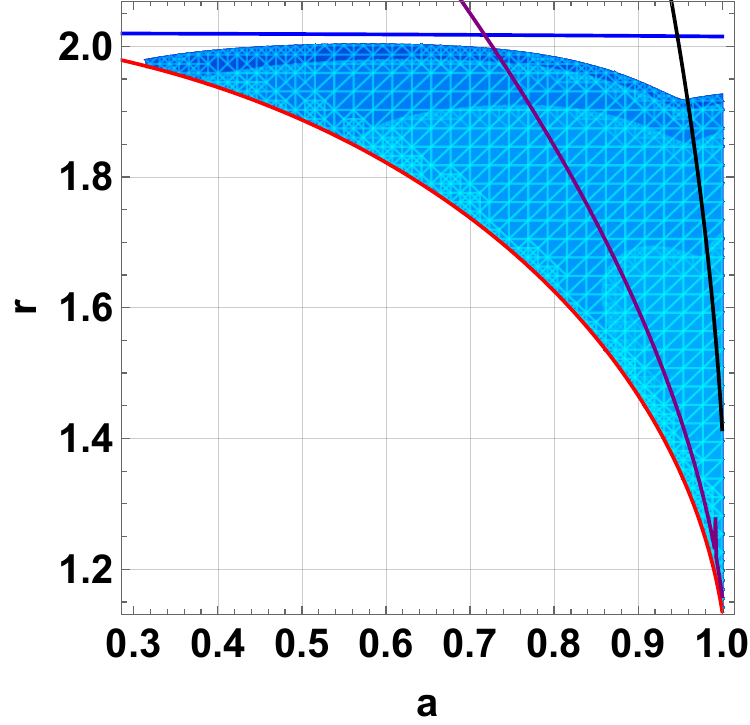}}
\subfigure[~$K=1/100$]{\includegraphics[width=4.7cm,height=4.3cm]{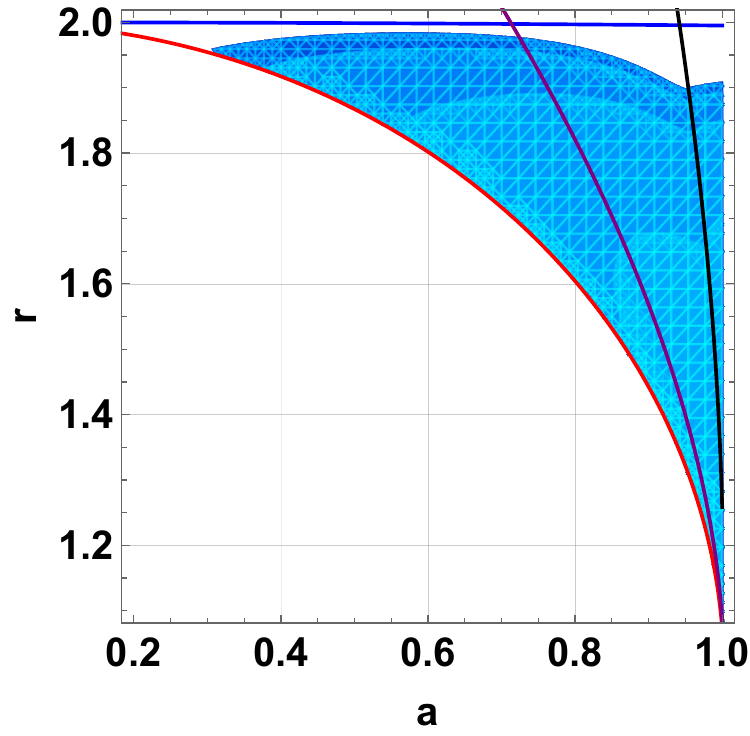}}
\caption{Plots showing the allowed energy extraction regions for different values of $K$, with fixed $l=20,~N=2$ and $\xi=\pi/12$ in the plunging region.}\label{fig22}
\end{center}
\end{figure}

\begin{figure}[H]
\begin{center}
\subfigure[~$\xi=\pi/12 $]
{\includegraphics[width=4.7cm,height=4.3cm]{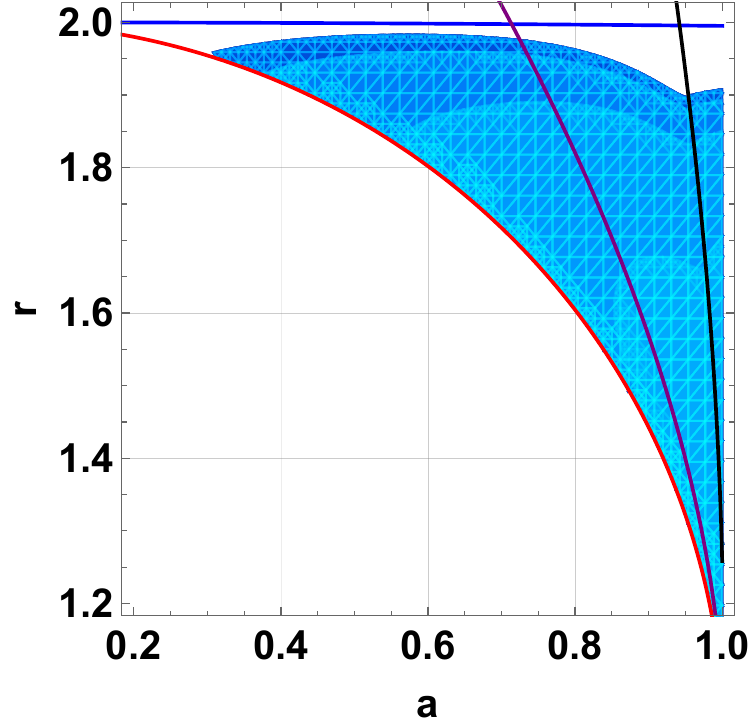}}
\subfigure[~$\xi=\pi/6
$]{\includegraphics[width=4.7cm,height=4.3cm]{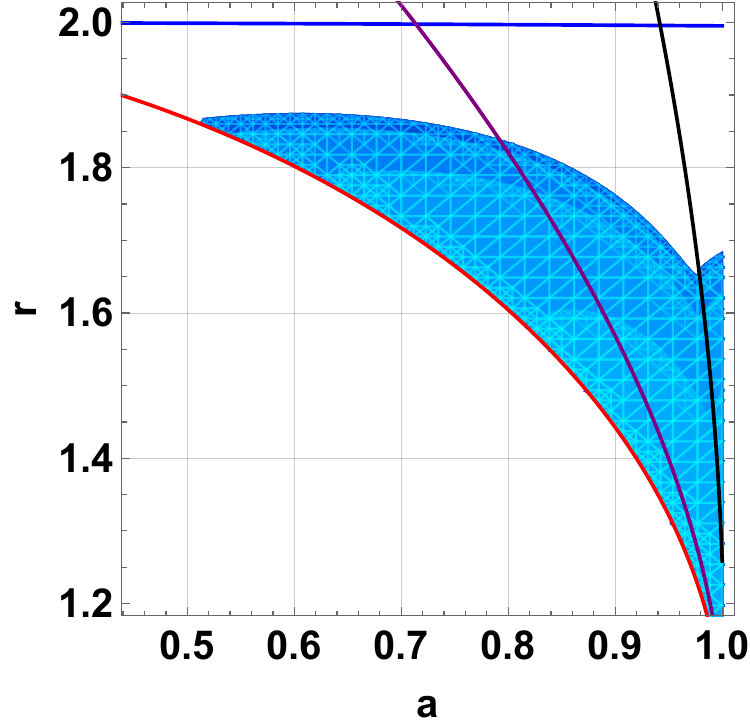}}
\subfigure[~$\xi=0
$]{\includegraphics[width=4.7cm,height=4.3cm]{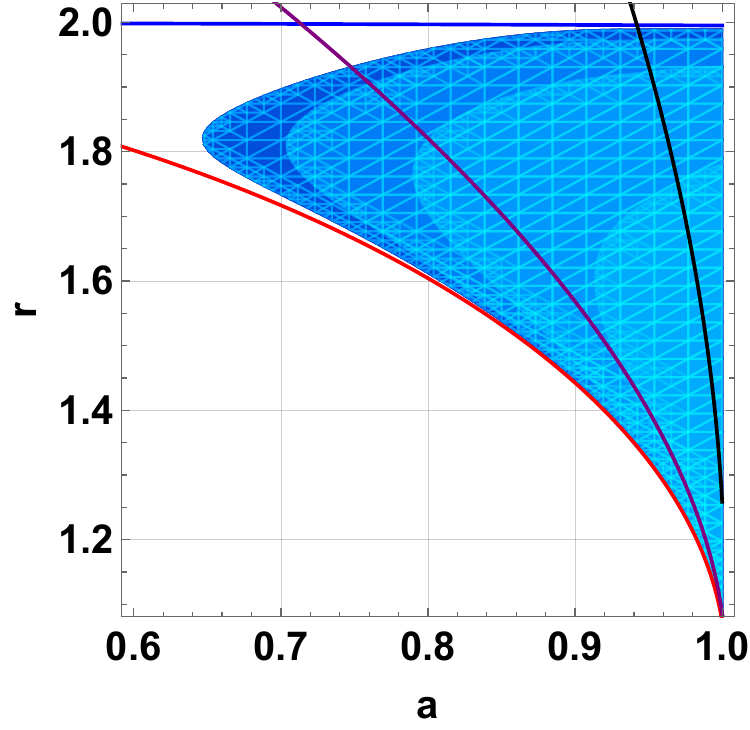}}
\caption{Plots showing the allowed energy extraction regions for different $\xi$, with fixed $l=20,~N=2$ and $K=1/100$ in the plunging region.}\label{fig23}
\end{center}
\end{figure}

In Fig.~\ref{fig22}, we demonstrates the allowed energy extraction regions for different values of $K$. We observe that as $K$ decrease, the minimum allowed spin for energy extraction in plunging
region is also decrease. While the maximum allowed spin shows similar behavior as we observed in circular orbits scenario. Also, in this case, the minimum allowed spin for energy extraction in
plunging region is smaller than the circular orbits region. As an example, if we take $\sigma=100$ and $K=1/20$, the minimum allowed spin in plunging region (Fig. \ref{fig22}(a)) is $0.3$ and in
circular orbits region (Fig. \ref{fig8newk}(c)) is $0.8$. From Fig.~\ref{fig23}, we notice that the shape of the allowed energy extraction region varies with the azimuthal angle. This behavior is a characteristic feature of the plunging region, where the dynamics explicitly depend on the azimuthal angle. These results are consistent with \cite{shen2024energy}.

\begin{figure}[H]\centering
\includegraphics[width=6cm,height=5.6cm]{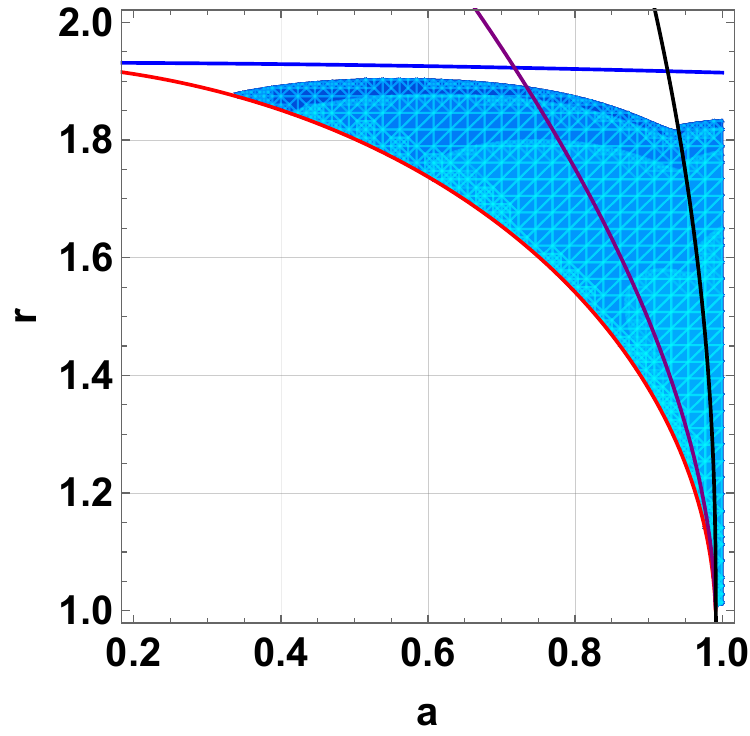}
\caption{Plot showing the allowed energy extraction region with fixed $K=1/500,~N=2,~l=10$ and $\xi=\pi/12$ in the plunging region.}\label{fig24}
\end{figure}
Following Fig.~\ref{fig10}, we exhibit the allowed energy extraction region for small values of $K$ and $l$ in Fig.~\ref{fig24}. We also observe that the minimum allowed spin is $0.2$.

\subsection{Power of Energy Extraction in the Plunging Region}

Based on Eq.~(\ref{powerratio}), we display the energy extraction power for some specific values of parameters in Fig.~\ref{fig25}. Specifically, we highlight the difference between the plunging and circular orbit cases within $r<r_I$. We consider $a=0.9,~l=17,~K=1/50,~N=2,~\xi=\pi/12$ and $\sigma=100$, here all values of $r$ are less than ISCO.

\begin{figure}[H]\centering
\includegraphics[width=6cm,height=5.6cm]{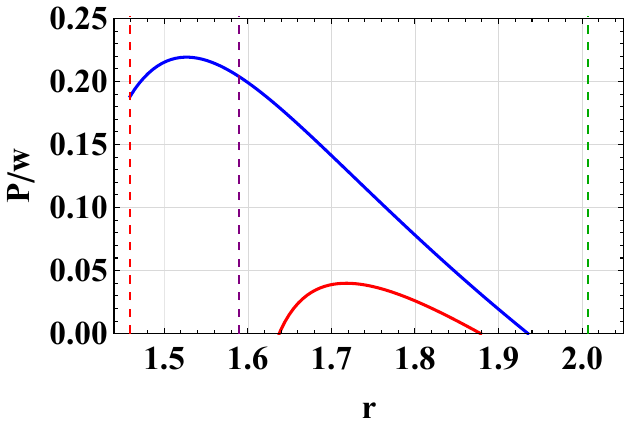}
\caption{Blue solid line represents the plunging region power and red solid line represents the circular orbit power. Green dashed, red dashed and purple dashed lines corresponds to the ergosphere, event horizon, and radius of photon sphere, respectively.}\label{fig25}
\end{figure}
From Fig. \ref{fig25}, we notice that the energy extraction power in plunging region is higher than that in the circular orbit region. Finally, we compare the power ratio for the plunging region with the Blandford-Znajek process using Eq.~(\ref{powerratio}). For desired results, we use the same values of relevant parameters as consider in the circular orbit case, namely $l=20,~K=1/100,~N=2,~\sigma=7, a=0.98$ and $r=1.4$. We found that $r_I=1.997>r$, which satisfies
the condition for the plunging region. Here, we obtain a power ratio of $6.744$, which is greater than $5.143$ obtained in the circular orbit case. This result demonstrates that the energy extraction power in the plunging region exceeds both the circular orbit power and the Blandford-Znajek power. Now considering a low spin scenario based on Fig. \ref{fig14}(c) ($N=1,~K=1/100,~l=20,~\xi=\pi/12$), spin $a$ can be $0.2$. However, in this case the radius of photon sphere is higher than the ergosphere, so Eq. \ref{crossArea} is no longer valid and should be modified as
\begin{equation}
A_{in}\approx (r^2_E-r^2_+).\label{newcrossarea}
\end{equation}
Taking $\sigma=100,~a=0.2,~r=2$ and $\xi=\pi/12$, we obtain a power ratio of $1.62912$, which is very small. This can be understood from the fact that, in the plunging region, obtaining energy extraction at low spin values need a relatively large value of $\sigma$. According to (\ref{powerratio}), $\sigma$ appears in the denominator, reducing the power ratio. Nevertheless, the possibility of extracting energy through magnetic reconnection at spin values as low as $0.2$ is specifically remarkable, since previous studies have rarely achieved energy extraction at such low spins.

\begin{center}
\section{CONCLUSION}
\end{center}

The investigating of magnetic reconnection illustrates a novel mechanism for extracting energy from rotating black holes. The efficient and feasible extraction of energy from black holes has
remained a significant topic among scientists in modern astrophysics. Motivated by this objective, in this work we investigate the energy extraction process in the context of rotating Einstein-AdS-SU($N$)-NLSM (or RASN) black hole model. The foundation of present work is the Comisso-Asenjo process \cite{comisso2021magnetic}. Both the plunging and circularly flowing bulk plasmas are considered for comparison. For desired results, we first discussed the physical behavior of event horizon, boundary of ergosphere, and the radius of the photon sphere. We notice that, due
to the influence of the coupling constant $K$, AdS radius $l$ and the flavors number $N$, results in decrease the maximum allowed spin as compared to the Kerr black hole. Then, we discuss the mechanism for the energy extraction in circular orbit with the help of magnetic reconnection. We interpret the physical behavior for different values of parameters and analyzed how values of $K,~N,~l, \sigma$ and $\xi$ affect on energy extraction. We also analyzed the power and efficiency of energy extraction, and particularly studied the energy extraction power and efficiency under the parameters $l,~K$ and $N$.

Moreover, we compared it with the Blandford-Znajek mechanism and concluded that magnetic reconnection power is greater than that of the Blandford-Znajek mechanism power. Finally, we investigated the energy extraction process in the plunging region and discussed the underlying magnetic reconnection mechanism. We also analyzed the influence of the parameters $K,~N,~l,~\sigma$ and $\xi$, and compared the obtained results with those of the circular orbit
scenario. Our result shows that both the power and efficiency of energy extraction are significantly higher in the plunging region. For circular orbits, we observe that energy extraction is possible even for a spin parameter as low as $0.70$, whereas in the plunging region, energy extraction can occur for spins as low as $0.2$. This behavior appears due to the influence of the parameters $l,~K$ and $N$ related with the considering black hole model. We further interpret that these parameters facilitate energy extraction by reducing the minimum spin required in both the circular orbit and plunging region scenarios.

This conclusion shows that the power of energy extraction is highest in the plunging region. Our result suggests that working in the plunging region is most effective for energy extraction. For
circular orbits, decreasing $N$ and $K$ or increasing $l$ enables energy extraction at lower spin. Same pattern followed in plunging region that decreasing $N$ and $K$ or increasing $l$ enables energy extraction at lower spin. These results is entirely consistent with the findings \cite{comisso2021magnetic}. In summary, the present analysis reveals the intricate properties of energy extraction from RASN black hole model, deepening our understanding of black hole energy release mechanisms within this spacetime and providing a theoretical foundation for future observational studies of black holes. Moreover, investigations of energy extraction in other
rotating black hole models, along with more realistic astrophysical environments involving accretion flows \cite{camilloni2025self}, could offer further insights into energy extraction mechanisms.\\
\\
\textbf{Conflicts of Interest:} The authors declare that they have no conflicts of interest.\\
\\
\textbf{Data Availability Statement:} This study is purely theoretical and does not involve any associated datasets. Hence, data sharing is not applicable.

\end{document}